\documentclass[pre,floatfix,longbibliography,superscriptaddress,onecolumn,final,notitlepage]{revtex4-1}
\usepackage[utf8]{inputenc}
\usepackage[T1]{fontenc}

\usepackage{multirow}
\usepackage{subcaption}
\usepackage{tabularx}
\usepackage{graphicx,caption}
\usepackage{tikz}
\usepackage{pgfplots}
\usepackage{amsfonts,amsmath,amssymb}
\usepackage[mathcal]{euscript}
\usepackage{comment}
\usepackage{lineno}
\usepackage{hyperref}
\usepackage[explicit]{titlesec}
\usepackage{cleveref}
\usepackage{adjustbox}

\newcommand{\be}{\begin{equation}}
\newcommand{\ee}{\end{equation}}
\newcommand{\bea}{\begin{eqnarray}}
\newcommand{\eea}{\end{eqnarray}}

\renewcommand{\L}{\mathcal{L}}

\newcommand{\bern}{\textrm{Be}}

\newcommand{\pois}{\textrm{Pois}}

\newcommand{\ra}{\rightarrow}
\newcommand{\f}[2]{\frac{#1}{#2}}

\newcommand{\ccup}[1]{\left\{#1\right\}}

\newcommand{\bup}[1]{\left(#1\right)}
\newcommand{\rup}[1]{\left[#1\right]}

\newcommand{\s}{\sigma}

\newcommand{\Exp}{\mathbb{E}}

\renewcommand{\ref}[1]{[\ref{#1}]}

\usepackage{tabularx}

\usepackage{amsfonts,amsmath,amssymb}   
\usepackage{comment}
\usepackage{lineno}
\usepackage{hyperref}
\usepackage[explicit]{titlesec}

\usepackage{graphicx}
\usepackage[labelfont=bf]{caption}
\usepackage[format=hang]{subcaption}

\usepackage{algorithm,algorithmic}

\hypersetup{
 bookmarks=true, 
 unicode=false, 
 pdftoolbar=true, 
 pdfmenubar=true, 
 pdffitwindow=false, 
 pdfstartview={FitH}, 
 pdftitle={}, 
 pdfauthor={}, 
 pdfcreator={}, 
 pdfproducer={}, 
 pdfkeywords={} {} {}, 
 pdfnewwindow=true, 
 colorlinks=true, 
 linkcolor=red, 
 citecolor=blue, 
 filecolor=magenta, 
 urlcolor=cyan 
}
\usepackage[algoruled,algo2e]{algorithm2e}
\usepackage{sidecap}
\usepackage{boldline}
\usepackage{setspace}

\definecolor{darkgreen}{rgb}{0.0, 0.5, 0.0}

\newcommand{\sr}{\mbox{{\small \textsc{SpringRank }}}}
\newcommand{\xor}{\mbox{{\small \textsc{Xor }}}}
\newcommand{\mt}{\mbox{{\small \textsc{MultiTensor }}}}
\graphicspath{{./figures/}} 

\begin{document}

\title{The interplay between ranking and communities in networks}

\author{Laura Iacovissi}
\email{laura.iacovissi@uni-tuebingen.de}
\affiliation{Max-Planck Institute for Intelligent Systems, Cyber Valley, Tübingen 72076, Germany}
\affiliation{Bosch Industry on Campus Lab, University of Tübingen}

\author{Caterina De Bacco}
\email{caterina.debacco@tuebingen.mpg.de}
\affiliation{Max-Planck Institute for Intelligent Systems, Cyber Valley, Tübingen 72076, Germany}

\begin{abstract}
\textsl{Community detection and hierarchy extraction are usually thought of as separate inference tasks on networks. Considering only one of the two when studying real-world data can be an oversimplification. In this work, we present a generative model based on an interplay between community and hierarchical structures. It assumes that each node has a preference in the interaction mechanism and nodes with the same preference are more likely to interact, while heterogeneous interactions are still allowed. The sparsity of the network is exploited for implementing a more efficient algorithm. We demonstrate our method on synthetic and real-world data and compare performance with two standard approaches for community detection and ranking extraction. We find that the algorithm accurately retrieves the overall node's preference in different scenarios, and we show that it can distinguish small subsets of nodes that behave differently than the majority. As a consequence, the model can recognize whether a network has an overall preferred interaction mechanism. This is relevant in situations where there is no clear ``a priori'' information about what structure explains the observed network datasets well. Our model allows practitioners to learn this automatically from the data.}
\end{abstract}

\maketitle

\thispagestyle{empty}

\section{Introduction}

In interacting systems, an observed tie between two individuals can often be explained by the existence of groups or a hierarchical organization. For instance, in social networks interactions between people can be explained by the communities the individuals belong to \cite{10.2307/2577271,wasserman_faust_1994}. In animal interaction networks, animals fight or mate strategically based on some underlying notion of ranking perceived between them to determine their dominance in a hierarchy \cite{10.1093/comnet/cnab001}. When modelling network datasets, one typically observes only the set of interactions, but communities and ranking are hidden variables that need to be learned from the data, problems referred to as community detection and ranking extraction.
Often, practitioners consider only one of the two for the dataset at hand, expecting a clear contribution for either community structure or hierarchy in determining edge formation. However, there can be situations where this distinction is blurry, and it is not clear what among these two effects plays a bigger role in explaining the data.  For instance, students may report friendship relationships based on the groups they belong to or based on some hidden notion of hierarchy between them, thus reporting what friendship they aspire to have instead. The problem is that in these two cases the input dataset, a directed and possibly weighted network, looks the same. It is a list of edges $i \ra j$ and their weight $w$, but one may not know what mechanism best explains the observed data. Unless a practitioner has a strong \textit{a priori} expectation about how the system works, it is not clear how to distinguish if the network was generated by community structure or by hierarchical organization. The question thus is how to learn this interplay between communities and hierarchies from the data in an automatic way.

A large variety of algorithms are available for extracting communities and ranking from networks, spanning from simple heuristics and deterministic approaches to probabilistic ones. Standard algorithms for ranking entities are based on spectral methods, e.g.~PageRank \cite{ilprints422} and Eigenvector Centrality \cite{10.2307/2780000}. They are based on random walks on network and output real-valued scores. A different family of approaches considers ordinal rankings. These are typically extracted by finding an optimal permutation of the nodes that minimizes some penalty function. Relevant examples are Minimum Violation Rank \cite{10.2307/2631621, 52b32fcbb71d4ab7ac6739bec25c3399}, SerialRan \cite{NIPS2014_f0e52b27} and SyncRank \cite{cucuringu2016sync}. Other approaches that consider real-valued scores extracted from pairwise preferences are those based on Random Utility Models \cite{train2009discrete}, such as the Bradley-Terry-Luce (BTL) model \cite{bradley1952rank,luce1959possible}.  A different approach is that of \sr \cite{de2018physical}, a physically inspired ranking algorithm that computes real-valued scores via minimizing the energy of a system of springs representing the directed observed interactions.
In terms of community detection models, there are various traditional approaches like graph partitioning, spectral clustering modularity-based algorithms or divisive algorithms \cite{fortunato2010community}. In this work we focus on those based on probabilistic generative models \cite{goldenberg2010,bickel2009nonparametric,ball2011efficient,de2017community} like the Stochastic Block Model (SBM) \cite{HOLLAND1983109} and its variants. These have several advantages, including the possibility of sampling synthetic networks with a given community structure and predicting  missing links. Most importantly, they allow for a probabilistic treatment, which is the approach we adopt here to tackle our problem.

In most cases, these algorithms are applied independently, i.e.~one either extracts communities or ranking, and the practitioners decides a priori which model is most appropriate.
There have been some attempts to consider both communities and ranking as hidden variables on networks, assuming some underlying interaction between the two mechanisms. For instance, Chen at al. \cite{chen2015clustering} combine clustering and ranking by first inferring non-overlapping groups using a variant of SBM and then retrieving the within-cluster popularity of nodes of different types within each group, which in turns influences the community they belong to. Here the assumption is that nodes belong to groups and there is a ranking of nodes within each group. A recent work has studied the ranking communities problem, addressing the detection of communities by ranking them using information flow techniques \cite{10.1007/978-3-030-89912-7_22}. An additional way of mixing the concepts of community detection and hierarchy is hierarchical clustering. The main idea is that there exists a hierarchy of communities that can be organized via a tree structure \cite{10.1007/978-3-540-73133-7_1, DBLP:journals/corr/abs-2009-07196}. Finally, we mention clustering algorithms used for ranking data, i.e. data that are intrinsically embedded with a hierarchical structure given as metadata \cite{JACQUES2014201}.
All of these settings are fundamentally different from the problem considered here, in that they assume an intrinsic rank of communities or an intrinsic clustering of ranks. In this manuscript, instead, we assume that nodes interact mainly either because of community affinity or ranking, and we want our algorithm to learn the preferred mechanisms of each node automatically from the data.
This is a relevant problem in networks where the two mechanisms coexist. For example, an individual can have a contact with another individual because of homophily or because of prestige. The former case can occur when there are some common attributes and preferences, so that individuals recognize them as part of the same market. The latter corresponds to cases in which the individuals are perceiving themselves as part of some league in terms of prestige, and they aim at connecting with someone in the same league or slightly above \cite{doi:10.1126/sciadv.aap9815}.
In these systems, models considering just one mechanism will be likely to recognize a subset of the interactions as noisy observations, or even interpret them in their own terms, leading to a distorted interpretation of the underlying patterns.

To address this problem, in this work we propose a probabilistic model capable of recognizing the community and ranking structures in a network with coexisting mechanisms and quantifying the extent to which each individual prefers one instead of the other.  Our model considers latent variables encoding the division of nodes in clusters, the hierarchical organization of nodes and how every node prefers to interact. The generative model as we define it, allows us to address also the problems of predicting missing links in the data and assigning a preferred interaction type to each node, i.e. community affinity or ranking similarity.
We validate our model on synthetic data and showcase its applications on three real datasets where the impact of community and ranking mechanisms differ. We find that it is capable of correctly assigning to each node its preferred interaction mechanism (hierarchy or community) with high confidence, i.e. all the probabilities are close to zero or one. In addition, also the coefficient representing the overall preferred mechanism is always close to the ground-truth value. All of this is achieved without losing accuracy on the edge prediction task, whose performances on the limit cases are close to the ones of the baseline methods.

\section{Generating networks with coexisting community and hierarchical structures}

In this work, we are interested in modelling networks with underlying coexisting community and hierarchical structures. We refer to the network mainly through its matrix representation, i.e. the adjacency matrix $A = \{A_{ij}\}_{i,j=1}^N$, where $N$ is the number of nodes. The entry $A_{ij} \in \mathbb{N}$ represents the number of directed interactions $i \rightarrow j$ from node $i$ to node $j$. Each interaction can be either due to affinity between nodes (community) or competition between them (hierarchy). We are interested in scenarios where these two mechanisms coexist, and the interaction type is not known in advance. The goal is thus to observe a network and distinguish edges based on which of these two mechanisms is more likely to explain the interaction. To this end, we assume nodes to be of two types: those that predominantly interact through community, and those that predominantly interact through hierarchy. The intuition is that nodes with the same preference are more likely to interact, inducing the edge type (community or hierarchy). We further assume that the probability of a heterogeneous interaction between nodes of two different types is not null and can be considered as a third edge type. From a generative modelling perspective, this scenario can be understood as first drawing latent labels on nodes, corresponding to node types. Then drawing interactions between nodes from a specific distribution depending on their types, with their mean parameterized accordingly. Formally, the generative model is:
\bea
 \sigma_{i}, \sigma_{j} &\sim& \bern(\mu) \label{eqn:sigma}\\
A_{ij} &\sim& \begin{cases}  \pois(A_{ij};S_{ij})^{\s_i } \; \pois(A_{ij};M_{ij})^{1-\s_i} & \text{if} \quad \sigma_{i}=\sigma_{j} \quad \text{(\textit{in-group interaction})} \\ \pois(A_{ij};\delta_0) & \text{if} \quad\s_{i}\neq\s_{j} \quad \text{(\textit{out-group interaction})} \end{cases} \quad, \label{eqn:A}
\eea
where $\s_{i} \in \ccup{0,1}$ represents the node type, $\mu \in \rup{0,1}$ its prior, and $\delta_0 \geq0$ is a parameter that controls the density of edges between nodes of different type (typically small). The parameters $M_{ij}$ and $S_{ij}$ determine the community and hierarchy mechanisms, respectively. The procedure is repeated for each edge $i \rightarrow j$, since we assume conditional independence of the $A_{ij}$ given the latent random variables.
The model of Eqs. (\ref{eqn:sigma}), (\ref{eqn:A}) leads to the following network likelihood distribution
\be \label{eq:gen-model}
P(A | M,S, \s, \delta_0) = \prod_{ij}  \rup{ \pois(A_{ij};S_{ij})^{\s_i} \; \pois(A_{ij};M_{ij})^{1-\s_i} }^{\delta_{\s_i\s_j}} \, \rup{ \pois(A_{ij};\delta_0) }^{1-\delta_{\s_i\s_j}} \;.
\ee
Notice that with this parameterization we obtain that an edge type random variable $\delta_{\s_i\s_j} \in \ccup{0,1}$ can be naturally defined in terms of $\s$ using $\delta_{\s_i\s_j}=2 \s_{i}\s_{j}-\s_{i}-\s_{j}+1$. It is Bernoulli distributed with parameter $\mu^2 + (1-\mu)^2 = 1 - 2\mu(1-\mu)$.

To model community interactions parametrized by $M_{ij}$ we use \mt\cite{de2017community} (MT), a mixed-membership variant of the SBM.
Each node of the network belongs to a community to an extent represented by two membership vectors: $u_i = [u_{ik}]$ determines how much $i$ belongs to the community $k$ considering the amount of \textit{out-going} edges; $v_i = [v_{ik}]$ only considers \textit{in-coming} edges. An affinity matrix $w = [w_{kh}]$ encodes the density of edges between nodes in different communities. Note that all these quantities are positive but not necessarily normalized. These elements are combined in the expected number of community interactions as $M_{ij} = \sum_{k,h=1}^K u_{ik} v_{jh} w_{kh}$. This definition results in interactions more likely to exist between nodes with compatible community structure.

To model the hierarchical interactions parameterized by $S_{ij}$ we use \sr\cite{de2018physical} (SR), a model that associates a score $s_{i} \in \mathbb{R}$ to each node and an interaction energy $ \frac{\beta}{2}(s_i - s_j - 1)^2$ to each edge $i \rightarrow j$ which regulates the probability of a hierarchical interaction as a Boltzmann weight. Here $\beta$ is a hyperparameter that controls the strength of the hierarchy. These elements are combined in the expected number of hierarchical interactions as $S_{ij} = c \exp\rup{-\frac{\beta }{2}(s_i - s_j - 1)^2}$, where $c$ controls for network sparsity. This definition results in interactions more likely to exist between nodes with similar scores, i.e. close in rank.

We refer to our model as  \xor. This is parameterized by $\theta=(u,v,w,s,c,\delta_0,\mu)$ and it models the network likelihood distribution $P(A|\theta,\sigma)$ as in \Cref{eq:gen-model}. A graphical representation of the generative model is provided in \Cref{fig:genmodel}, together with a toy example of graph realization.
Notice that for $\s$ equal to a null or a unitary vector \xor reduces to  \mt or to \sr, respectively.

\definecolor{darkblue}{rgb}{0, 0.24, 0.64}
\definecolor{darkorange}{rgb}{1, 0.45, 0.0}
\definecolor{yellowgreen}{rgb}{0.6, 0.8, 0.4}

\begin{figure}[t]
  \captionsetup[subfigure]{justification=centering}
  \begin{subfigure}{.48\textwidth}
    \centering
    \begin{tikzpicture}[scale=1.3]
      \draw[draw=darkorange, fill=darkorange!30](-1.5,0.5) circle [radius=0.3] node {$u_i$};
      \draw[draw=darkorange, fill=darkorange!30](-1.5,-0.5) circle [radius=0.3] node {$v_j$};
      \draw[draw=darkgreen, fill=yellowgreen!40](1.5,0.5) circle [radius=0.3] node {$s_i$};
      \draw[draw=darkgreen, fill=yellowgreen!40](1.5,-0.5) circle [radius=0.3] node {$s_j$};

      \draw(0,0) circle [radius=0.3] node {$A_{ij}$};

      \draw[draw=darkorange, fill=darkorange!30](-1.5,-2) circle [radius=0.3] node {$w$};
      \draw[draw=darkblue, fill=darkblue!20](0,-2) circle [radius=0.3] node {$\delta_0$};
      \draw[draw=darkgreen, fill=yellowgreen!40](1.5,-2) circle [radius=0.3] node {$c$};

      \draw [->] (-1.2,0.5) -- (-0.3, 0.1);
      \draw [->] (-1.2,-0.5) -- (-0.3, -0.1);
      \draw [->] (1.2,0.5) -- (0.3, 0.1);
      \draw [->] (1.2,-0.5) -- (0.3, -0.1);

      \draw [->] (-1.3,-1.75) -- (-0.2, -0.25);f
      \draw [->] (0,-1.7) -- (0, -0.3);
      \draw [->] (1.3,-1.75) -- (0.2, -0.25);

      \draw[rounded corners] (-2.2,1.7) rectangle (2.2,-1.25) {};

      \draw(0,1.2) node {$\forall \ (i, j) \in E$};

      \node[color=darkorange, rotate=+90] at (-2.5,0) {\textbf{$\s_i = \s_j = 0$}};
      \node[color=darkgreen, rotate=+90] at (2.5,0) {\textbf{$\s_i = \s_j = 1$}};
      \node[color=darkblue] at (0,-2.8) {\textbf{$\s_i \neq \s_j$}};
    \end{tikzpicture}
    \caption{Graphical model representation.}
    \label{fig:relations}
  \end{subfigure}
  \begin{subfigure}{.5\textwidth}
    \centering
    \begin{adjustbox}{width=0.6\linewidth}
    \begin{tikzpicture}[rotate=50, scale=1.5]

      \begin{scope}[every node/.style={color=black,fill=darkorange!50, circle,
                                      inner sep=6pt}]

        \node (a) at (-.5,4.8) {};
        \node (d) at (1.8,4.2) {};
        \node (c) at (-.3,2.8) {};
        \node (e) at (-2,1) {};
        \node (g) at (2,1.5) {};
      \end{scope}

      \begin{scope}[every node/.style={color=black,fill=yellowgreen!50, circle,
                                      inner sep=6pt}]

        \node (b) at (-2.5,3) {};
        \node (f) at (-0.8,0.8) {};
        \node (h) at (-.5,-1.5) {};
        \node (i) at (1.3,-.0) {};

      \end{scope}

      \begin{scope}[every edge/.style={draw=darkorange,very thick}]
        \path [-latex] (a) edge (c);
        \path [-latex] (c) edge (e);
        \path [-latex] (c) edge (d);
        \path [-latex] (d) edge (g);
        \path [-latex] (g) edge (c);
      \end{scope}

      \begin{scope}[every edge/.style={draw=yellowgreen,very thick}]
        \path [-latex] (f) edge (h);
        \path [-latex] (i) edge (h);
        \path [-latex] (i) edge (f);
      \end{scope}
      \begin{scope}[every edge/.style={draw=darkblue, thick}]
        \path [-latex] (c) edge (b);
        \path [-latex] (f) edge (c);
      \end{scope}

    \end{tikzpicture}
    \end{adjustbox}
    \vspace{0.2cm}
    \caption{Network example.}
    \label{fig:example}
  \end{subfigure}
  \caption{Model visualization. (a) Graphical model: the entry of the adjacency matrix $A_{ij}$ is determined by the community-related latent variables $u, v, w$ (orange), by the ranking-related ones $s, c$ (green) and by the out-group interaction parameter $\delta_0$ (blue), depending on the values taken by the node type latent variables $\sigma_{i},\sigma_{j}$. $E$ denotes the set of network directed edges. (b) Example of possible realization of the model: orange nodes interact mainly via community, i.e. $\s_{i}=0$, green ones via hierarchy, i.e. $\s_{i}=1$. Orange and green edges are interactions between nodes of the same type (matching node color), while  blue edges are interactions between nodes of different types. }
  \label{fig:genmodel}
\end{figure}
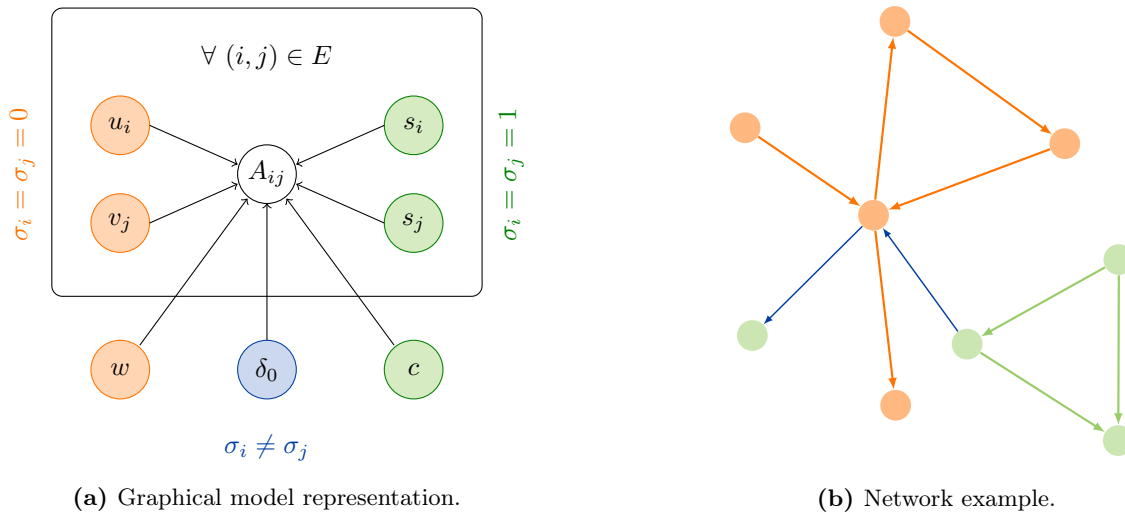

\section{Inference}

Given a network adjacency matrix $A$, we want to infer the parameters $\theta$ and the node labels $\sigma$ that best explain the observed data. To this end, we aim at maximizing $P(\theta| A) = \sum_{\s}P(\s,\theta| A)$, i.e. the maximum a posterior estimate of $\theta$. For convenience we maximize its logarithm instead, as the maxima coincide. We than take a variational approach by using Jensen's inequality
\be \label{eq:F}
\log P(\theta| A) = \log\sum_{\s}P(\s,\theta| A) \geq  \sum_{\s} q(\s) \log \f{P(\s,\theta| A)}{q(\s)} =: \L(q,\theta) \;,
\ee
where $q(\s)$ is a variational distribution over the node labels. This formulation of the problem turns it in a maximization of the function $\L(q,\theta)$ with respect to $\theta$ and $q$. In fact, since $q(\s)$ must sum to one, the exact equality between the second and third term in \Cref{eq:F} is achieved when $q(\s)$ is equal to the posterior $P(\s|\theta,A)$. However, this posterior may not be analytically accessible as the normalization is not tractable. In fact we have
\be \label{eq:postsigma}
P(\s|\theta,A) \propto P(\s, A | \theta) = \prod_i \mu^{\s_i}(1-\mu)^{1-\s_i} \prod_{ij} \, \pois(A_{ij};S_{ij})^{\s_i\delta_{\s_i\s_j}} \pois(A_{ij};M_{ij})^{(1-\s_i)\delta_{\s_i\s_j}} \, \pois(A_{ij};\delta_0)^{1-\delta_{\s_i\s_j}} \quad,
\ee
and this cannot be simply recast into a well-known probability distribution in $\s$ (e.g. a fully factorized Bernoulli distribution).
A careful reader will recognize that the $P(\s|\theta,A)$ is equivalent to an Ising model with unitary inverse temperature and Hamiltonian:
\begin{align}
  H_{A}(s|J,h) &= \sum_{i,j}J_{ij}s_{i}s_{j}+\sum_{i}h_{i}s_i\;,\label{eqn:isingH} \\
  J_{ij} &= \f{\log\pois(A_{ij};S_{ij})+\log\pois(A_{ij};M_{ij})-2\,\log\pois(A_{ij};\delta_0)}{4}\;,  \label{eqn:isingJ}\\
  \begin{split} \label{eqn:isingh}
    h_{i} &= \f{1}{4}\sum_{j}\bup{\log\pois(A_{ij};S_{ij})+\log\pois(A_{ji};S_{ji})-\log\pois(A_{ij};M_{ij}}-\log\pois(A_{ji};M_{ji})) \\ &\quad+ \f12 \bup{\log \mu - \log(1-\mu)}\;,
  \end{split}
\end{align}
where here $s_i \in \ccup{\pm 1}, \ s(\s)= 2\, \s-1$ and the couplings $J$ are asymmetric, see \Cref{S2}.

To obtain a tractable expression for the variational distribution that estimates $P(\s|\theta,A)$, we use a mean-field approximation $q(\s) = \prod_i q_i(\s_i)$ assuming $q_i(\s_i) = \bern(\s_i; Q_i)$. The goal is to find values of $\ccup{Q_{i}}_{i}$ such that the Kullback-Leibler divergence between the approximate posterior $q(\s)$ and the true posterior $P(\s|\theta,A)$ is minimized \cite{opper2001advanced,blei2017variational}. Noting that the Hamiltonian in Eq.~(\ref{eqn:isingH})  corresponds to $- \log P(\s,A | \theta)$, the maximization of $\L(q,\theta)$ with respect $q$ is equivalent to minimize a variational free energy $F(q,\theta)$ defined as:
\be \label{eqn:freeE}
  F(q,\theta ) = \sum_{\s} q(\s) \,H_{A}(s(\sigma),|J(\theta),h(\theta))  - S(q) = - \L(q,\theta) + \log P(\theta)\;,
\ee
with the first term being the internal energy of $q$ and $S$ the entropy function of the product of Bernoulli distributions $q(\s)$.

By performing this minimization for $q$ we obtain that the optimal parameters that are included in \Cref{alg:EM}, i.e. \Crefrange{eq:updateQ}{eq:updateQ2}, see \Cref{S3} for detailed derivations. This result can also be obtained using the standard self-consistency equation for an Ising model $\s_{i} = \tanh\bup{h_{i}+\sum_{j}J_{ij}\s_{i}}$ using $J_{ij}$ and $h_{i}$ as in
\Cref{eqn:isingJ,eqn:isingh}. One can in principle use alternative approximations more complex than mean-field \cite{opper2001advanced}, for instance the Bethe approximation, at the cost of increasing computational complexity. This is left for future work.

The values of $Q_{i}$ are also point-estimates for the variables $\s_{i}$, as for a Bernoulli distribution $\Exp_{q}[\s_{i}]=Q_{i}$ .
Differentiating $\L(q,\theta)$ with respect to $\theta$ and setting this to zero gives the updates for the parameters $\theta$. The full derivation is reported in \Cref{S4}, while the results can be seen in \Cref{alg:EM} where we show the overall EM algorithmic routine. The algorithm does not guarantee convergence to the global maximum of the variational log-likelihood, but only to a local one. In practice, we perform different runs with different random initializations of the inputs and select the one with the best value of the $\L(q,\theta)$.

The computational complexity per iteration scales as $O(E K^2 + N^2)$, where $E$  is the total number of directed edges. In most of the applications, $K$ is usually much smaller than $E$. For sparse networks, as is often the case for real datasets, $E \propto N$. Hence, we have a complexity that is dominated by $O(N^2)$. This contribution comes from terms containing  $\tilde{Q}_{ij} =Q_{i}Q_{j}$ that are not also multiplied by $A_{ij}$, i.e. terms in the denominators of the updates in \Cref{alg:EM}.  The matrix $\tilde{Q}=\rup{\tilde{Q}_{ij}}$ is  a dense object and was not present in the updates of \mt, whose computational complexity is $O(E\, K^{2})$, nor in that of the updates of \sr, whose complexity is that required to solve a sparse linear system. This may make it prohibitive to run our model on large systems. In these cases, one can consider approximating $\tilde{Q}$, e.g. by batch sampling of pairs $(i,j)$ as done in machine learning applications \cite{hoffman2013stochastic}. We do not explore this here.


\setlength{\textfloatsep}{5pt}
\begin{algorithm}[H]
\SetKwInOut{Input}{Input}
	\setstretch{0.7}
	\Input{ network $A=\{A_{ij}\}_{i,j=1}^{N}$; number of communities $K$; inverse temperature $\beta$; $w$'s, $u$'s and $v$'s a priori parameters $\lambda_u,\; \lambda_v, \; \lambda_w \in [0,1]$.}
  	\BlankLine
	\KwOut{membership vectors $u=\rup{u_{ik}},\, v=\rup{v_{ik}}$; network-affinity matrix $w=\rup{w_{kq}}$; ranking score vector $s=\rup{s_{i}}$; outgroup interaction parameter $\delta_0$; $\s$'s a priori parameter $\mu$;  $\s$'s a posteriori parameters' vector $Q=\rup{Q_{i}}$.}
	\BlankLine
	 Initialize $u,v,w,s,c,\delta_0,\mu,Q$ at random.
	 \BlankLine
	 Repeat until $\L$ converges:
	 \BlankLine
	\quad 1. Calculate $\rho$ and $Q$ (E-step): 
	\begin{align}
	  \rho_{ijkh} &= \frac{u_{ik}v_{jh}w_{kh}}{\sum_{lm}u_{il}v_{jm}w_{lm}} \;, \\
    Q_{i} &= \f{f_{i1}}{f_{i1} + f_{i2}} \;, \label{eq:updateQ} \\ 
    f_{i1} &= \mu \prod_{j\neq i}\rup{\pois(A_{ij};S_{ij})\pois(A_{ji};S_{ji})}^{Q^{\text{old}}_{j}}\rup{\pois(A_{ij};\lambda_{0})\pois(A_{ji};\lambda_{0})}^{(1-2Q^{\text{old}}_{j})} \;, \label{eq:updateQ1} \\
    f_{i2} &= (1-\mu) \prod_{j\neq i}\rup{\pois(A_{ij};M_{ij})\pois(A_{ji};M_{ji})}^{Q^{\text{old}}_{j}-1} \label{eq:updateQ2} \;;
	\end{align}

	 \quad 2. Update parameters $\theta$ (M-step):
	\BlankLine
	\quad \quad \quad
		i) for each node $i$ and community $k$ update memberships:
		\be
		\quad  u_{ik}= \f{\sum_{jh} (Q_iQ_j -Q_i - Q_j +1)\, A_{ij}\rho_{ijkh} }{\lambda_u + \sum_{jh}(Q_iQ_j -Q_i - Q_j +1)\, v_{jh}w_{kh} } \;, \quad  v_{ik}= \f{\sum_{jh} (Q_iQ_j -Q_i - Q_j +1)\, A_{ij}\rho_{ijkh} }{\lambda_v + \sum_{jh}(Q_iQ_j -Q_i - Q_j +1)\, u_{jh}w_{kh} } \;;  \label{eq:updateuv}
		\ee
	\quad \quad \quad
	ii) for each pair $(k,h)$ update affinity matrix:
		\be
		\quad w_{kh} =\f{\sum_{ij} (Q_iQ_j -Q_i - Q_j +1)\, A_{ij}\rho_{ijkh} }{\lambda_w +\sum_{ij}(Q_iQ_j -Q_i - Q_j +1)\, u_{ik}v_{jh}} \;; \label{eq:updatew}
		\ee
  \quad \quad \quad
    iii) update the sparsity coefficient and the ranking scores for each node $i$:
    \be
    \quad c = \f{\sum_{ij} Q_{i}Q_{j}\, A_{ij}}{\sum_{ij}Q_{i}Q_{j} \exp \rup{-\f{\beta}{2} (s_{i}-s_{j}-1)^{2}}}\;, \quad s_i = \f{\sum_{j} Q_{i}Q_{j} \,s_j \rup{A_{ji} + A_{ij}  } + Q_{i}Q_{j} \rup{ A_{ij} - A_{ji} }}{\sum_j Q_{i}Q_{j} \rup{A_{ji} + A_{ij}}} \;;  
    \ee
  \quad \quad \quad
		iv) update outgroup interaction parameter:
    \be
    \quad \delta_0 = \f{\sum_{ij}A_{ij} (2Q_iQ_j -Q_i - Q_j) }{\sum_{ij} (2Q_iQ_j -Q_i - Q_j) } \;.  
    \ee
	\quad \quad \quad
		v) update $\mu$ parameter:
    \be
    \quad \mu = \f{\sum_{i} Q_i}N \;.  
    \ee
	\caption{\xor:  EM algorithm}
	\label{alg:EM}
\end{algorithm}

In the model we did not specify any prior for the $\theta$. An alternative is to impose exponential priors for each $u_{ik}, v_{ik}$, independent and identically distributed with parameters $\lambda_u, \lambda_v$. This results in a $L_{1}$-regularized $\L(q,\theta)$ that enforces sparse membership vectors and an overall contribution of the community mechanism close to $||w||_{max}$. To control the growth of the value $||w||_{max}$ we can impose an exponential prior on $w$ as well, parametrized by $\lambda_w$. The regularizer may prevent the SBM to overfit. This might be particularly relevant in the case of a block structure induced on the network by different leagues in the hierarchical organization of the nodes. The prior results in small modifications of the updates for the $u,v,w$, which are reported in equations \Cref{eq:updateuv,eq:updatew}. We do not set a prior on $s$ since in previous studies \cite{de2018physical} it was shown that adding a Gaussian prior may not necessarily lead to better prediction performance.

\section{Results}
\subsection{Results on synthetic data}

\begin{figure}[t]
  \captionsetup[subfigure]{justification=centering}
  \begin{subfigure}{.32\textwidth}
    \centering
    \includegraphics[scale=0.55]{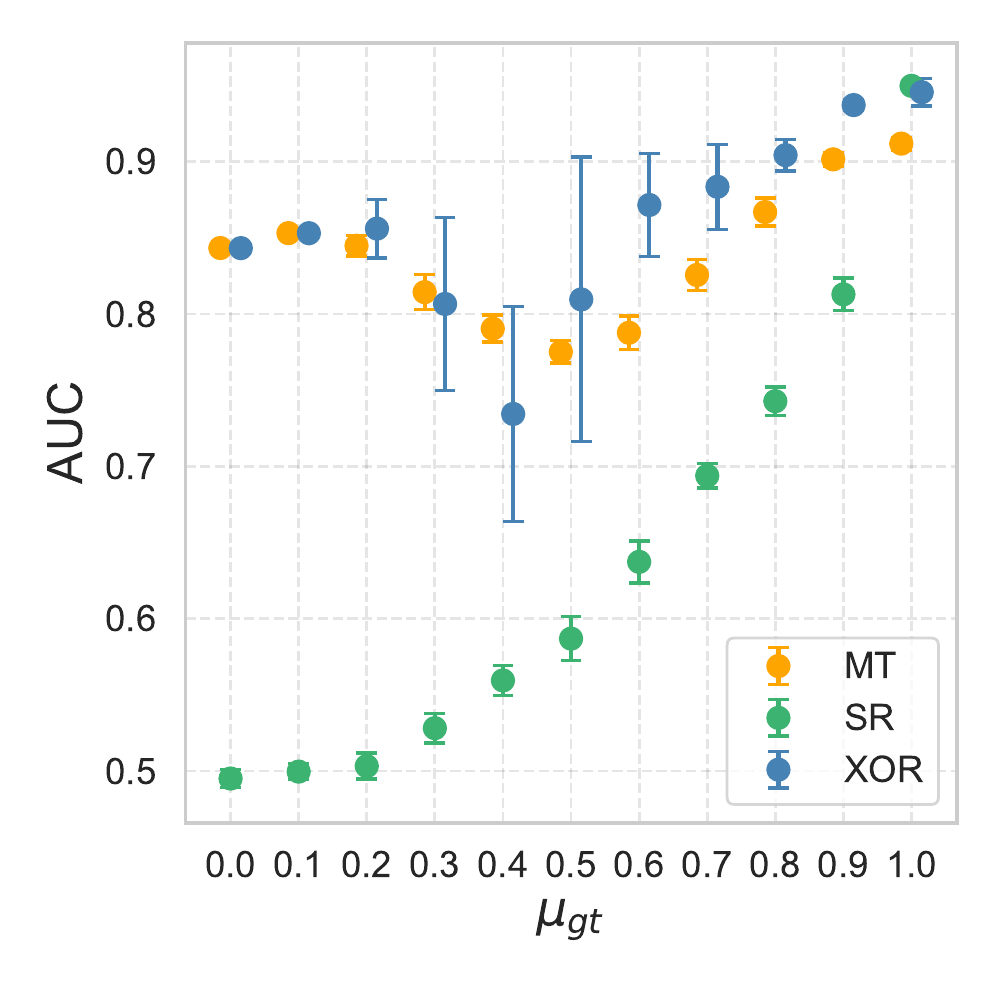}
    \label{fig:edgepred}
    \caption{Edge prediction task.}
  \end{subfigure}
  \begin{subfigure}{.32\textwidth}
    \centering
    \includegraphics[scale=0.55]{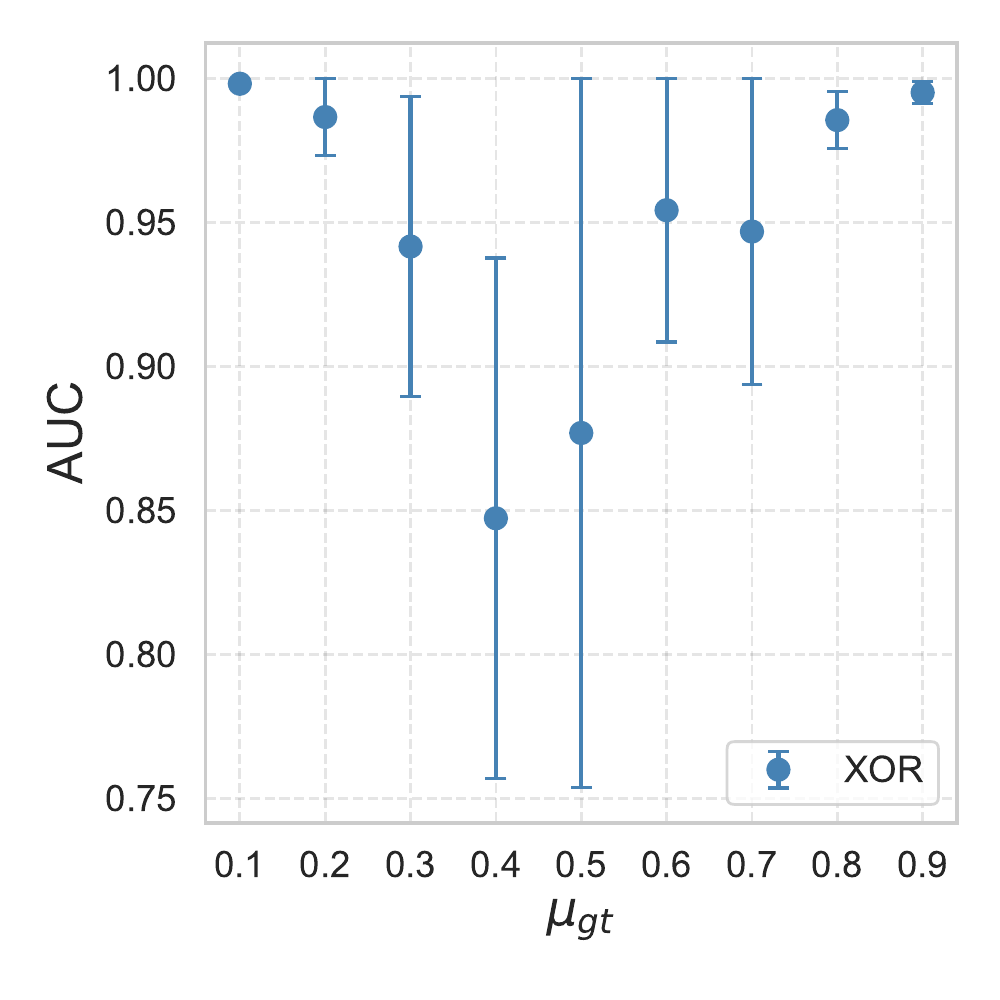}
    \label{fig:typepred}
    \caption{Node type prediction task.}
  \end{subfigure}
  \begin{subfigure}{.32\textwidth}
    \centering
    \includegraphics[scale=0.55]{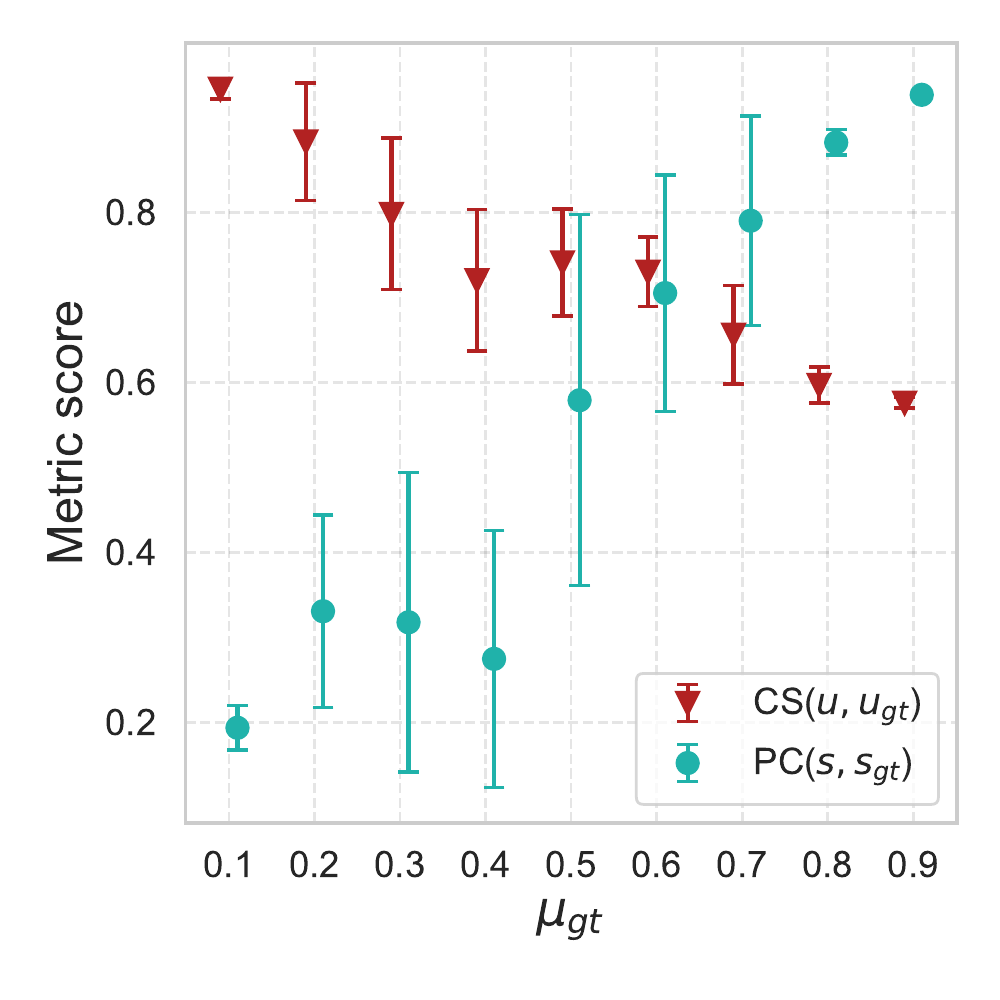}
    \label{fig:infmetrics}
    \caption{Latent variables' inference task.}
  \end{subfigure}
  \caption{Performances on synthetic networks for different tasks. The mean value across folds is reported, indicating also the standard deviations. For the edge prediction task, our model is compared with the baseline methods \mt and \sr. We vary the proportion of expected nodes with $\s_i =1$, i.e.~preferring a hierarchy-based interaction. Only the not regularised version of \xor is shown since it is the most effective in this scenario.}
  \label{fig:syntAUC}
\end{figure}

The \xor model outputs the parameters related to the community $(u,v,w)$, the hierarchical structure $(s,c)$ and the node types $\s$ from the observed network data $A$. When the ground-truth values of $\theta$ and $\s$ are available, we can measure the performance of the model in recovering the community structure, the ranking of nodes and their type. For these three tasks we consider as performance metrics the cosine similarity (CS), the Pearson's correlation (PC) and Area Under the Curve (AUC), respectively, between the ground-truth and the inferred values.  In the absence of ground-truth, we can indirectly evaluate the model fitness via edge prediction tasks in cross-validation settings where we hide a subset of the matrix $A$ (test set), fit the model in the remaining subset (training set) and test the ability to predict the missing edges (test set).

We validate the model on synthetic data generated using the \xor generative model with $N=500,$ average degree $\langle k \rangle=20$, $\beta=5$  and varying the ground-truth value of $\mu_{gt} \in [0,1]$. Specifically, we generate networks with $K=3$ communities of equal-size unmixed group membership, a hierarchy with $l=3$ leagues, i.e. the scores $\ccup{s_{i}}_{i=1}^{N}$ are drawn from a mixture of Gaussians with means $\{ -4, 0, 4\}$ and standard deviations $\{1, 0.5, 1\}$; we set $\delta_0=0.01$. We draw five different independent samples for each set of parameters.
The inference algorithm is tested by using 5-fold cross-validation for splitting the data into train and test sets, and ran with five different random initializations on each graph instance. We set the hyperparameters $K$ and $\beta$ equal to the ground-truth values, while we use grid search for selecting the best value of the regularization $\lambda = \lambda_v = \lambda_u = \lambda_w \times 0.1$.

We find that \xor predicts missing edges robustly and consistently across different values of $\mu_{gt}$ and better than baseline models that consider only community structure (MT) or only hierarchical structure (SR), see \Cref{fig:syntAUC} (left). The performance is not monotonic in $\mu_{gt}$: we obtain high values of AUC when $\mu_{gt}$ is close to $0$ or when $\mu_{gt}$ is close to 1. These are extreme scenarios where a large majority of the nodes are predominantly interacting either via communities  or hierarchy. As one of these two mechanisms dominates, it is also easier to infer the parameters, hence the higher AUC values. The intermediate region $0.2 \leq \mu_{gt}\leq 0.8$ where the nodes distribute more evenly between the two mechanisms corresponds to cases in which inference is harder. Nevertheless, the model shows stable performance in this range, with AUC mean values never dropping below 0.7 and always comparable or better than MT, the best performing among the two baselines. A similar non-monotonic behaviour is observed for the node classification task, where we aim at predicting the node type $\s$ using $Q$.
While we observe a similar performance drop in the same intermediate regime, performance is robust, as the average AUC is always higher than 0.85.
Finally, the performance in recovering communities is good in the regime that most favours community structure ($\mu_{gt}<0.4$) while recovering of the hierarchy is poor, and vice-versa in the opposite regime ($\mu_{gt}>0.6$). This is intuitive, as when most of the nodes predominantly interact via communities, their score is irrelevant, and therefore cannot be recovered well. What matters is the ability to recover the structure corresponding to the main mechanism at play, i.e. high cosine similarity when $\mu_{gt}<0.5$ or high Pearson's coefficient when $\mu_{gt}>0.6$. We find that \xor performs well in this task, with a boost in performance in recovering the ranking to values above 0.7 in the regime where hierarchy dominates. Similarly, cosine similarity increases above 0.7 when communities dominate. These results suggest that the model is not only able to predict missing edges and nodes' type, but also to distinguish the node-level latent features that determine how nodes interact, regardless of their type.

\subsection{Results on real data}

\subsubsection*{Application on High School network}

\begin{figure}[t]
  \captionsetup[subfigure]{justification=centering}
  \begin{subfigure}{.27\textwidth}
    \centering
    \includegraphics[scale=0.38]{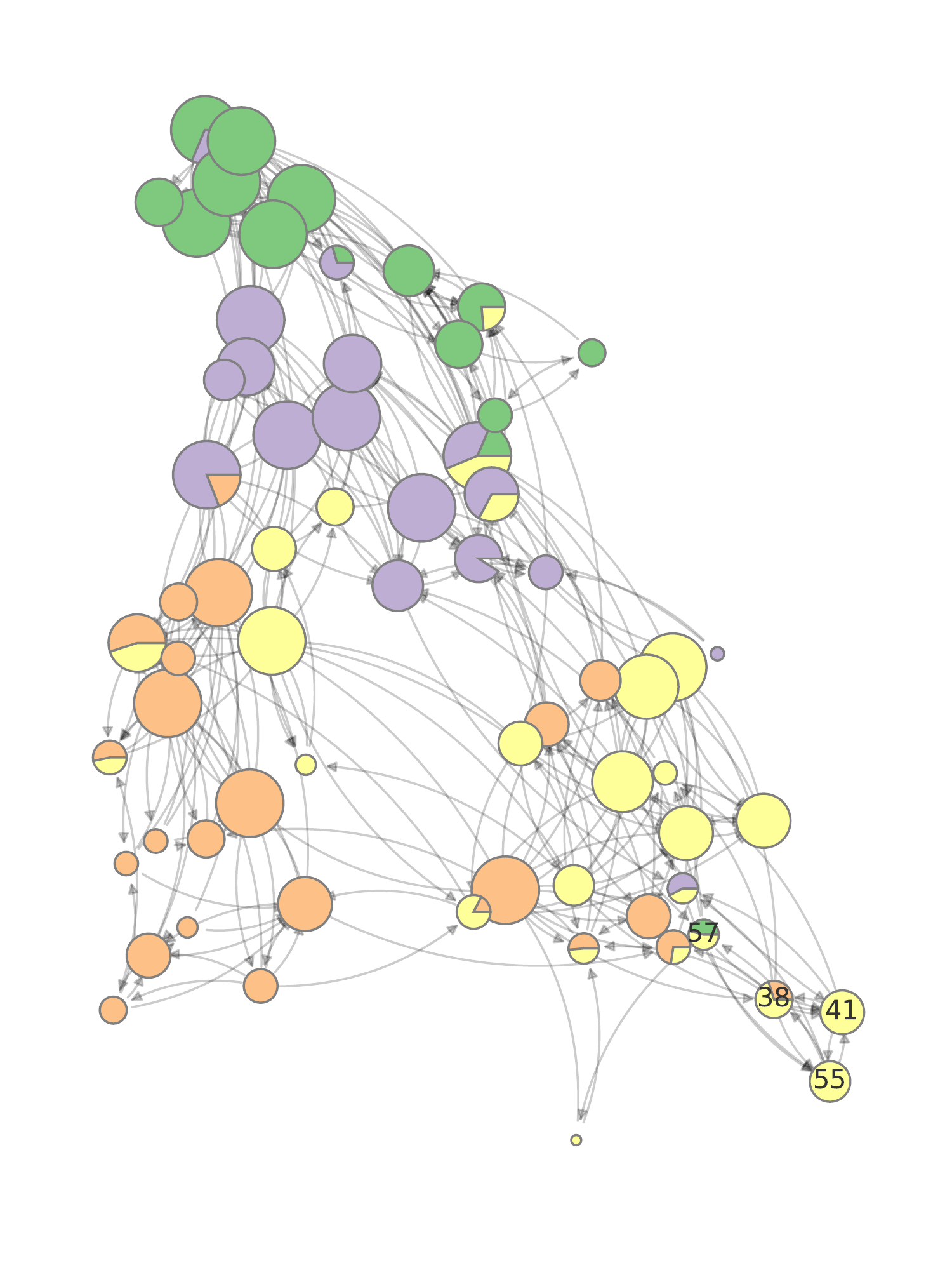}
    \label{fig:H-MT}
    \caption{\mt overlapping communities.}
  \end{subfigure}
  \begin{subfigure}{.27\textwidth}
    \centering
    \includegraphics[scale=0.38]{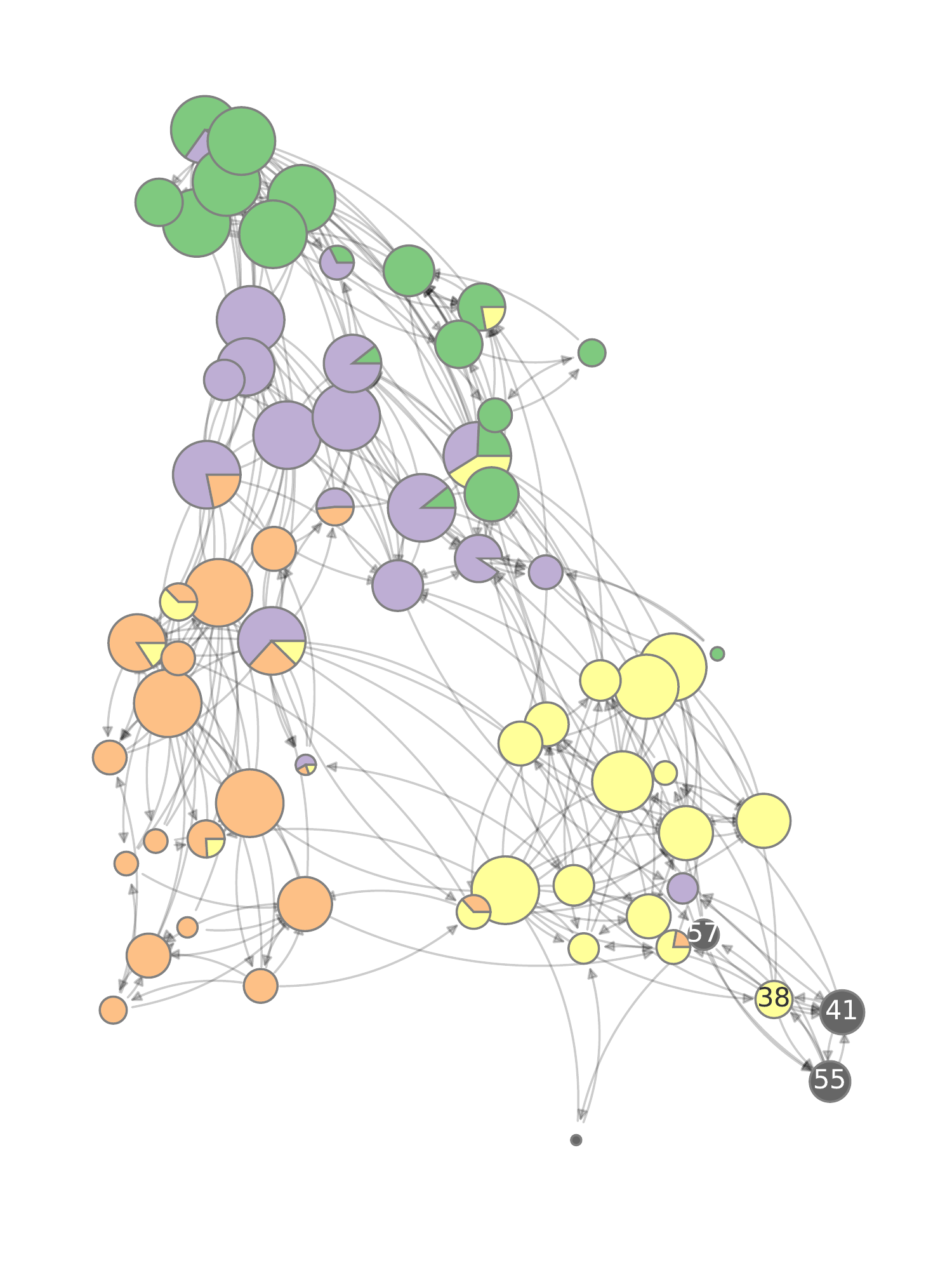}
    \label{fig:H-XOR}
    \caption{\xor overlapping communities.}
  \end{subfigure}
  \begin{subfigure}{.27\textwidth}
    \centering
    \includegraphics[scale=0.38]{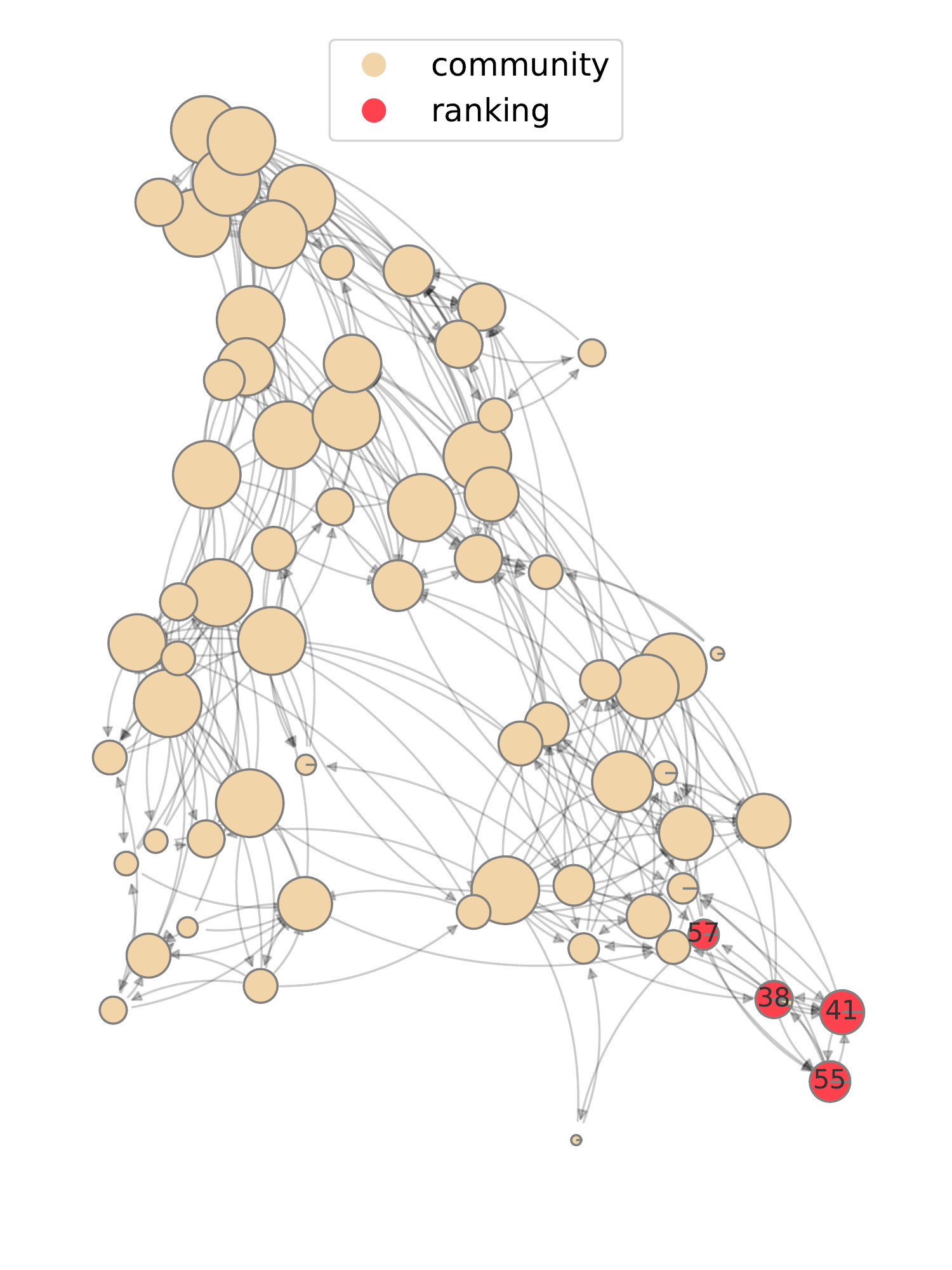}
    \label{fig:H-Q}
    \caption{\xor node types.\\ \hspace{2cm}}
  \end{subfigure}
  \captionsetup[subfigure]{justification=centering}
  \begin{subfigure}{.12\textwidth}
    \centering
    \includegraphics[scale=0.38]{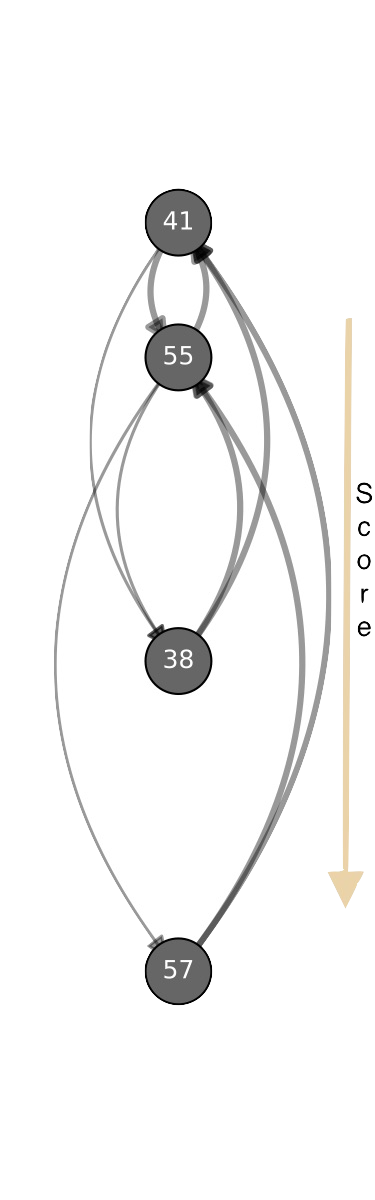}
    \label{fig:H-rank}
    \caption{\xor subnetwork hierarchy.}
  \end{subfigure}
  \caption{ Application on High School network: (a)-(b) comparison between the communities detected by \mt  and \xor, (c) node types $Q$ visualization and (d) hierarchy in the subnetwork of nodes preferring the competitive mechanism. Nodes' positions are assigned: (a)-(c) using a spring layout and (d) using the scores $s$ inferred by the \xor algorithm. Communities are selected by normalizing the $v$ membership vector (similar results are obtained with $u$) and colors of the pie markers are assigned according to the community (mixed) membership in each group. The dark grey nodes have null membership vector. Both the algorithms select $K=4$ as optimal number of communities, as output using a five-fold cross-validation scheme and grid search.}
  \label{fig:highschool}
\end{figure}

As a first example of application of this model, we consider a dataset of a network of high school students \cite{konect}. This describes the perceived interactions of a group of 67 high school students. Each student is being asked ``who are you friend of'' in the fall of 1957 and the spring of 1958, and the answers are aggregated on the same edge allowing weights with values 1 or 2. Agreement in the response is not ensured, hence the network is directed. It is reasonable to expect that students belong to groups (communities) and this influences the answers they give. However, a fraction of the students may not belong to any group and instead nominate others based on their perceived ranking of the students. For instance, one may nominate whom they aspire to befriend. \Cref{fig:highschool} shows what happens when we apply \xor to this dataset. The figure shows the estimates of node type $\s_{i}$, describing the preferred behavior of each student. As we can see, most of the students have a $\s_1\approx 0$, their nominations follow a community structure. However, we obtain four individuals with high $\s_i$, meaning that their preferences are mainly based on ranking. Even though they have a degree similar to that of other students, they are not well connected with the rest of the network as they mostly interact among themselves (there are only 5 other students that nominate one of them as friends). Comparing the reciprocity coefficient for the whole set of students with the one for the subgraph interacting mainly via hierarchy, we find that it increases from a value of $0.51$ to a $0.88$. In fact, this smaller network is missing only three edges for being a (directed) clique. Considering the direction and weight of the edges of the subgraph made of these four individuals, the ranking is meaningful as it reveals insights on the group social dynamics, as shown in \Cref{fig:highschool}(d), where the hierarchy highlights consistency between inferred score (the position of the nodes) and the weight and direction of the interactions. Namely, the individual with highest ranking in that subset ($i=41$), is also the individual that makes fewer friendship nominations, while the others tend to nominate them more often.

In the same figure, we show the community structure inferred by our model and that using \mt, which considers only this mechanism for edge generation. As we can see, \xor  outputs slightly different communities, as the yellow and blue nodes are partially mixed in the two cases. What is interesting, is that most of the nodes that have $\s_i=1$ are not assigned to any community: again, we observe that the latent variable related to the mechanism not used for interacting is meaningless.
Our models thus were able to extract a subgraph where hierarchy structure could meaningfully explain the directed interactions within that subgraph and also distinguish this from the remaining part of the networks where community memberships had likewise a meaningful interpretation. This showcases how practitioners should consider $\s_i$ together with $\mu_i$ and $s_i$ to fully characterize nodes.

\subsubsection*{Application on Parakeets network}

\begin{figure}[t]
  \captionsetup[subfigure]{justification=centering}
  \begin{subfigure}{.3\textwidth}
    \centering
    \includegraphics[height=0.282\textheight]{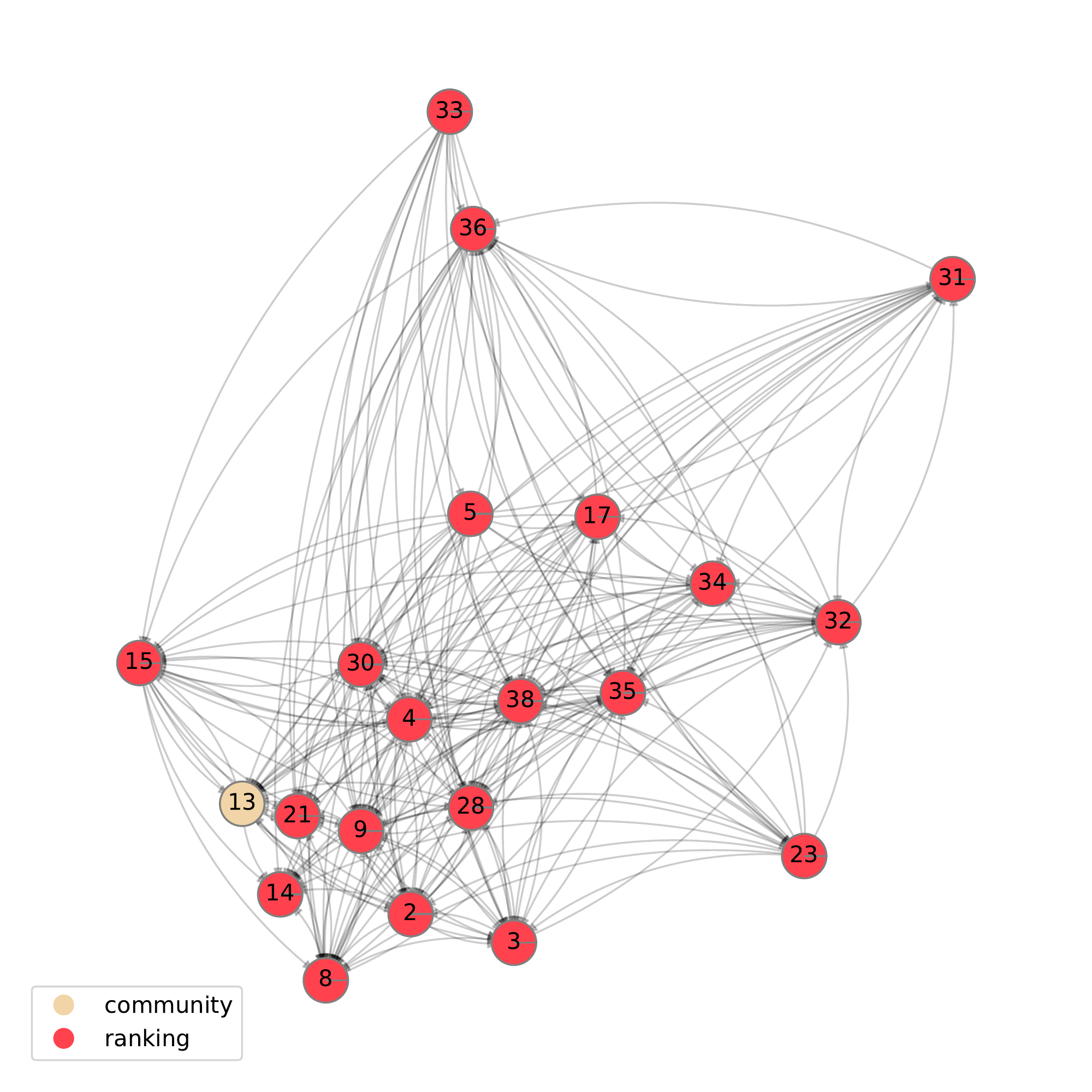}
    \label{fig:PG1-types}
    \caption{\xor node types.\\ \vspace{.9cm}}
  \end{subfigure}
  \begin{subfigure}{.3\textwidth}
    \centering
    \includegraphics[height=0.28\textheight]{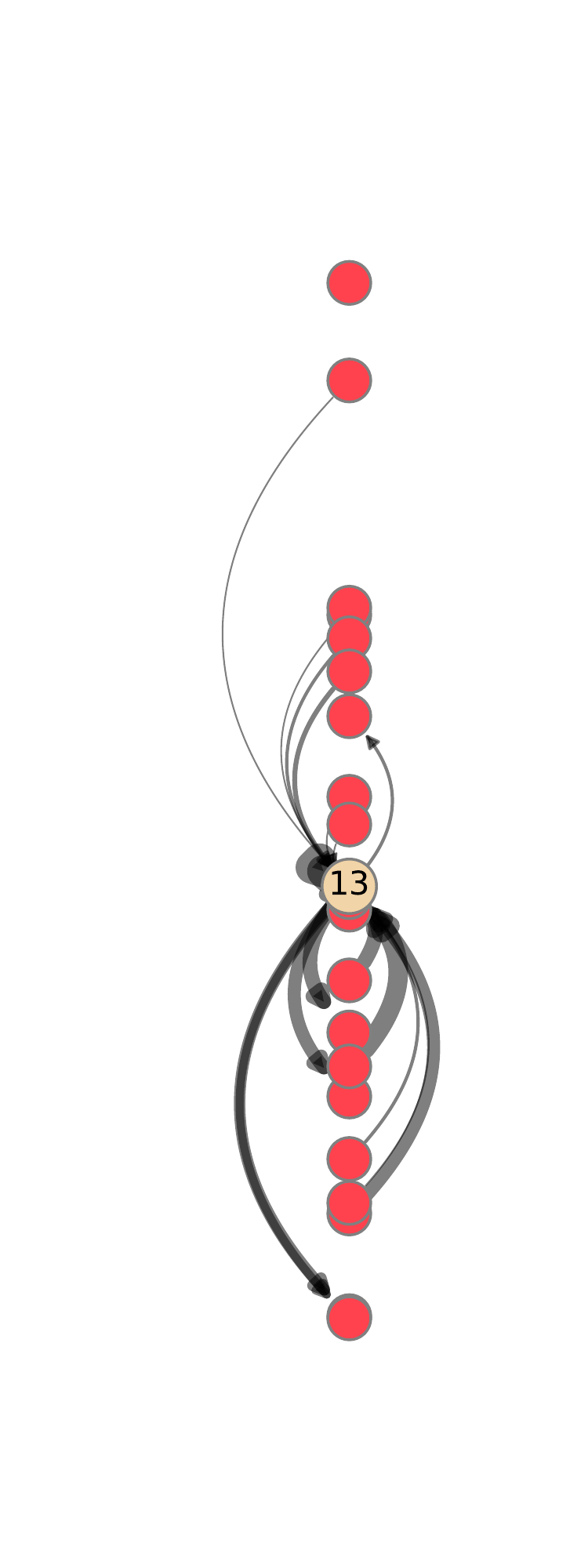}
    \label{fig:PG1-ranks}
    \caption{\xor ranking scores and edges involving node 13.}
  \end{subfigure}
  \begin{subfigure}{.3\textwidth}
    \centering
    \includegraphics[height=0.28\textheight]{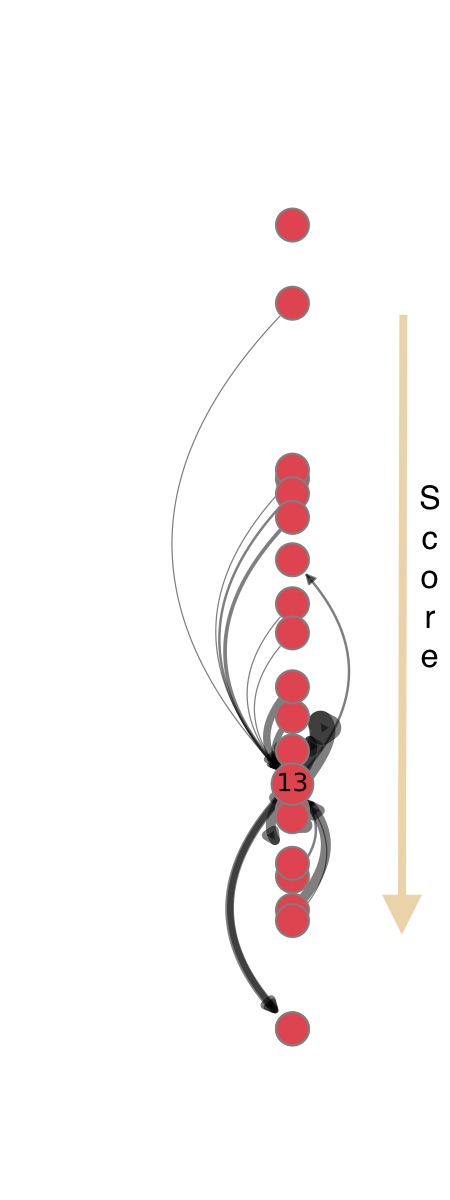}
    \label{fig:PG1-ranks}
    \caption{\sr ranking scores and edges involving node 13.}
  \end{subfigure}
  \caption{Application on network of parakeets, group 1: visualization of (a) the node types $Q$; (b) ranking scores, as inferred by the \xor model; (b) ranking scores inferred by the \sr model. In (b) and (c) we highlight the interactions involving node 13, the only one that \xor infers as interacting via communities, i.e. $\s_{13}=0$.  The node positions are (a) determined with force atlas and (b)-(c) based on the inferred score $s$ by \xor and \sr respectively. In (b)-(c), ranking scores are decreasing from top to bottom, as pointed out by the arrow on the right. Both \xor and \mt detect no meaningful community structure, since they select $K=1$ from five-fold cross validation and grid search.}
  \label{fig:parakeetsG1}
\end{figure}

A second case study is the application to a network of directed aggressions among captive monk parakeets (\textit{Myiopsitta monachus})  \cite{parakeets.dataset}. Each directed edge contained the number of aggressive attacks between parakeets in two study groups, the first made of 21 individuals (G1), the second of 19 (G2). Each group was created and then observed for 24 days, divided into four 6-day study quarters. Insights from behavioural ecology suggest that the patterns of aggression correlate with an underlying dominance hierarchy: parakeets direct their aggression strategically, aiming at improving their position in the hierarchy \cite{10.1371/journal.pcbi.1004411}. Hence, we expect them to have a prevalent hierarchical latent structure, i.e. we expect $\s_{i}=1$ for most of the individuals.

We performed experiments on the two groups, both extracting a single network from each quarter and aggregating on the quarters. The results on aggregated and not aggregated versions are similar for both groups, apart from more noisy results on the first quarter of both G1 and G2 given by a hierarchy not mature enough for being clearly detected \cite{10.1371/journal.pcbi.1004411}. The resulting inference on the aggregated G1 group is shown in  \Cref{fig:parakeetsG1}(a):  only one node is predicted to use community-based interactions. This is also reinforced by the results of cross-validation tests to extract the number of communities -- the value $K=1$ achieves the best AUC score -- and by inspecting $u, v$, which have mainly null entries. Because there is only one node detected with $\s_{i}=0$, the interpretation of community in this case is that this particular node has a behavior that cannot be well explained by the same mechanism that well explains that of all the other nodes (hierarchy structure in this case). We can deduct that this node is an anomaly in the social group, who interact with the other individuals with a random behaviour, rather than a strategic one. Hence, its score is to be considered irrelevant, which is in accordance with the fact that it is placed in a middle-ranking position while having a high number of incoming connections (33 incoming vs 40 outgoing, considering the weights), see \Cref{fig:parakeetsG1}(b). Note that the results achieved by \sr on the same network are similar: in \Cref{fig:parakeetsG1}(c) $i=13$ is placed 13th, close to the 10th position assigned by \xor, while the interaction pattern is not well in agreement with that. Again, it is behaving as an anomaly, but the \sr algorithm cannot learn it as it is not designed to distinguish node types.

\subsubsection*{Application on Political Blogs network}

\begin{figure}[t]
  \captionsetup[subfigure]{justification=centering}
  \begin{subfigure}{.27\textwidth}
    \centering
    \includegraphics[scale=0.2]{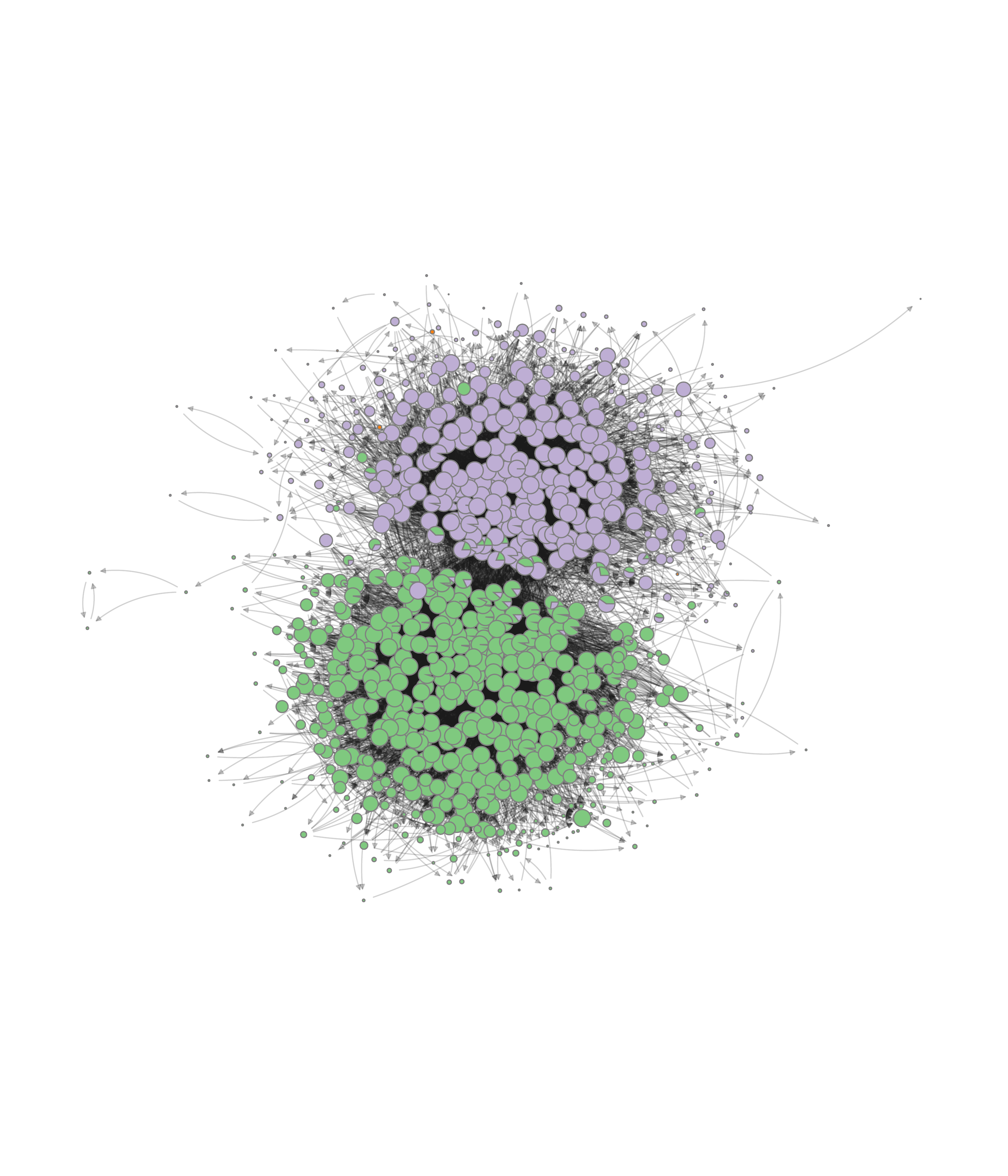}
    \label{fig:PB-MT}
    \caption{\mt overlapping communities.}
  \end{subfigure}
  \begin{subfigure}{.27\textwidth}
    \centering
    \includegraphics[scale=0.2]{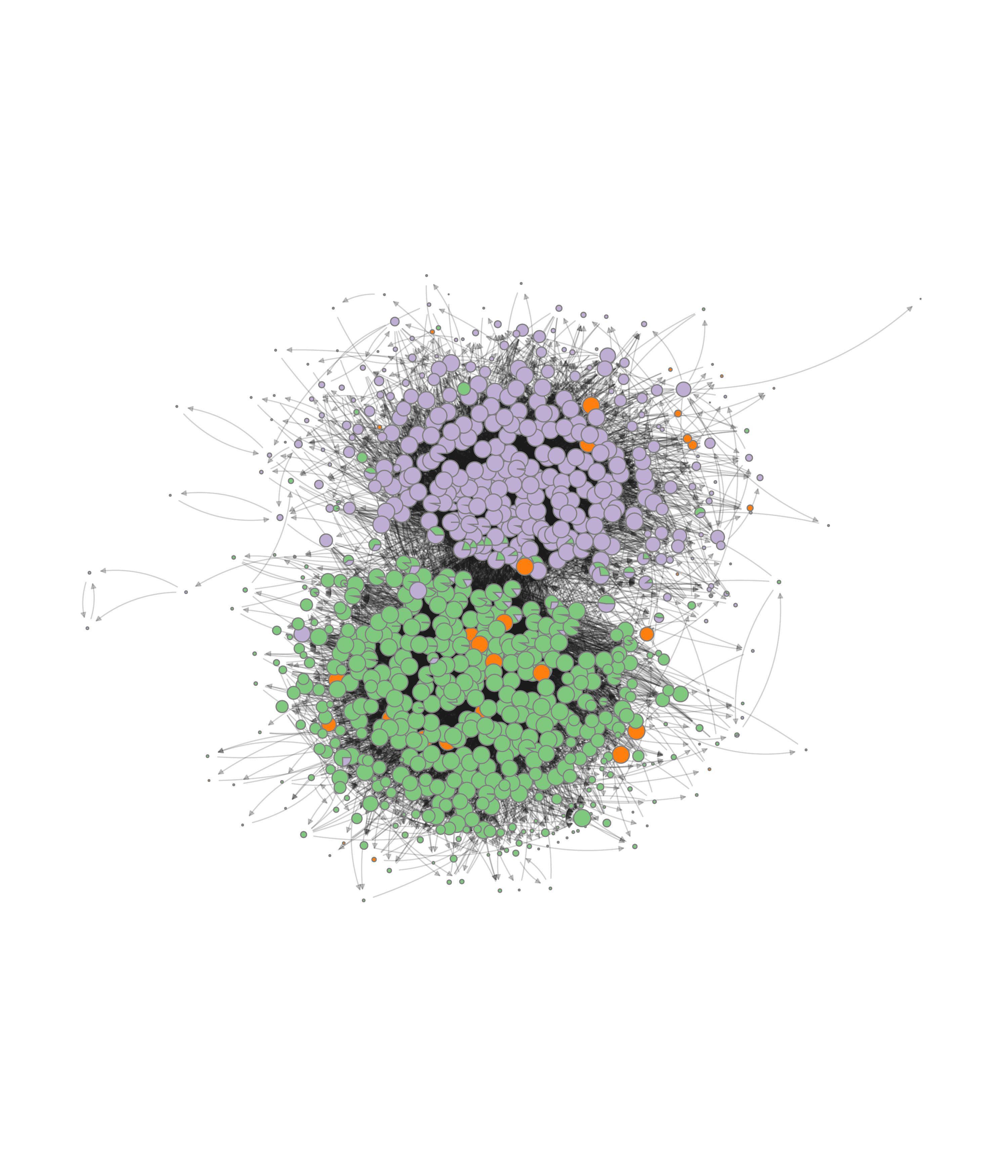}
    \label{fig:PB-XOR}
    \caption{\xor overlapping communities.}
  \end{subfigure}
  \begin{subfigure}{.267\textwidth}
    \centering
    \includegraphics[scale=0.2]{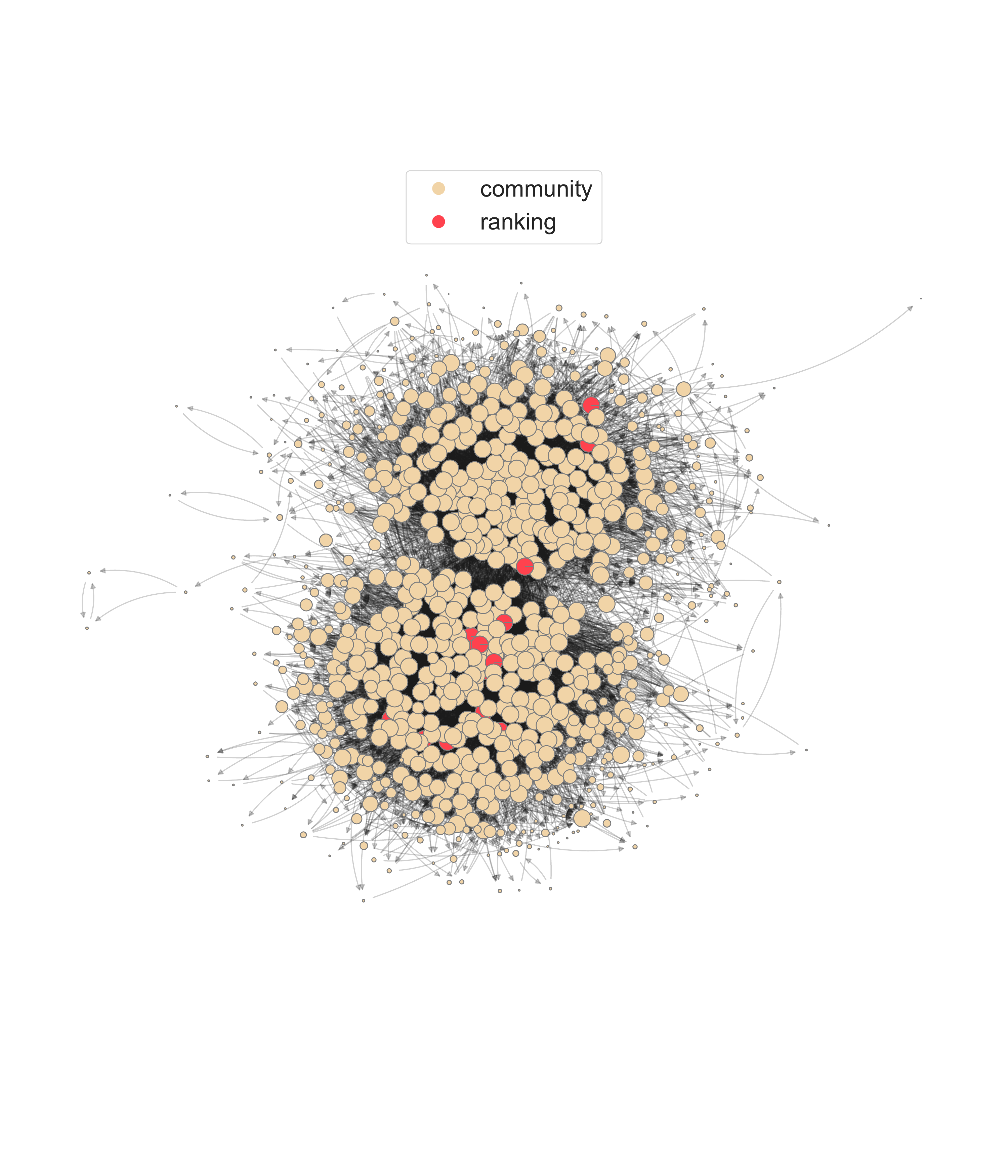}
    \label{fig:PB-Q}
    \caption{\xor node types.\\ \hspace{2cm}}
  \end{subfigure}
  \captionsetup[subfigure]{justification=centering}
  \begin{subfigure}{.15\textwidth}
    \centering
    \includegraphics[scale=0.054]{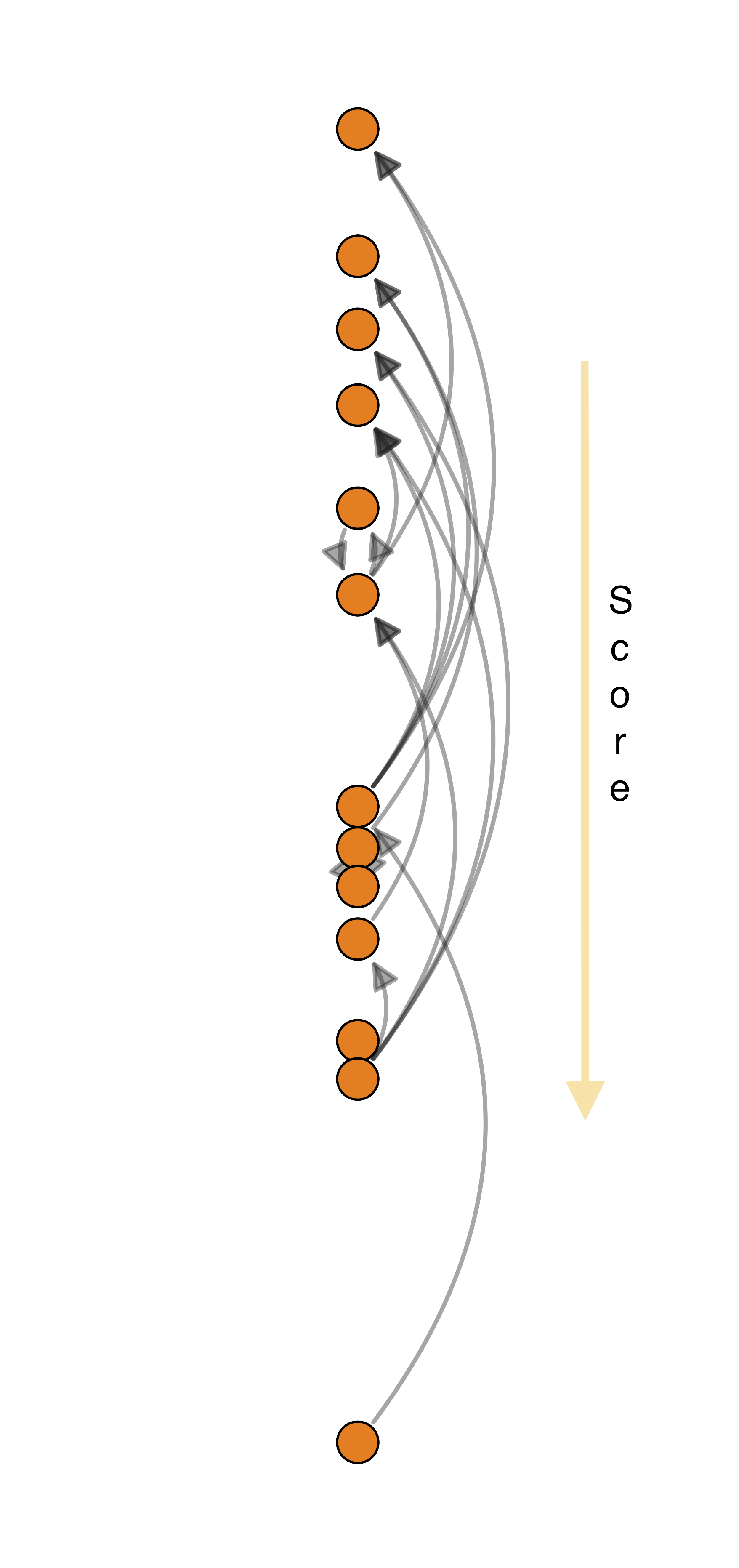}
    \label{fig:PB-rank}
    \caption{\xor subnetwork hierarchy.}
  \end{subfigure}
  \caption{Application on a network of political blogs. (a)-(b) comparison between the communities detected by \mt  and \xor, (c) node types $Q$ visualization and (d) hierarchy in the subnetwork of nodes preferring the competitive mechanism. Nodes' positions and communities are assigned as in \Cref{fig:highschool}. Both the algorithms select $K=2$ as optimal number of communities via $5$-fold cross-validation scheme and grid search. Here we show the in-coming membership $v$, orange nodes have null  $v_{ik}=0 \ \forall k$. \mt only assigns null in-coming community membership to nodes with null in-degree (5 nodes, not visible in the plot); \xor assigns null membership to these 5 nodes and to additional 32 ones (all plotted in orange). Of these 32 nodes, 13 are flagged as ranking-driven while 19 have low in-degree (less then 4). A similar behavior is observed considering $u_i$ and the out-going degree.}
  \label{fig:polblogs}
\end{figure}

As our final example we focus on the ability of \xor to recognize a dominant preferred interaction mechanism in larger datasets.
We consider a network of 830 US web blogs with different political orientations, where the weighted directed edges are the number of hyperlinks from a webpage to another. The data have been collected over the period of two months before of the 2004 US presidential election \cite{adamic2005political} and include metadata about each blog's political orientation -- liberal and conservative.
As shown in \Cref{fig:polblogs}, \xor identifies the large majority of the nodes ($98\%$) as driven by community structure. In addition, the algorithm selects $K=2$ as the best number of groups via cross-validation, in line with the observation that the nodes form two highly assortative groups; the assigned communities are in accordance with both those retrieved by \mt and the node metadata (\xor/MT total accuracy of $95\% / 95\%$; $97\% / 98\%$ on `liberal', $92\% / 91\%$ on `conservative'). These are also identified with a high confidence, as given by the low number of community overlapping.

The remaining small fraction of nodes, less than $2\%$, is classified as ranking-driven. The hierarchical structure between them is strong, as shown in \Cref{fig:polblogs}(d) where we notice only few edges violating the hierarchy, i.e., going from top to bottom. In this case, as in the High School network, the semantic of the directed edges is such that the most popular node is the one with the most in-coming edges.
In the absence of further information about these nodes, we investigated their network structural properties, finding no relevant feature that allows to distinguish them from the rest of the nodes. Their reciprocity, in- and out-degree and assortativity w.r.t. the two political orientations are distributed similarly as for the rest of the nodes. While we cannot rule out noise as a potential explanation for their distinct classification, we argue that this is a relevant example where domain knowledge may help explaining \textit{a posteriori} potential discordant patterns involving a small fraction of the nodes.
Overall, this example shows the robustness of \xor in identifying dominant preferred interaction mechanism in large networks and illustrates how practitioners can use these results to guide further investigations in the absence of a clear \textit{a priori} orientation towards communities or hierarchies.

\section{Discussion}

The \xor model captures coexisting hierarchical and community mechanisms in networks. Being a generative model, it can be used for producing synthetic benchmarks with the desired level of interplay between the two mechanisms. It relies on a principled  mathematical formulation with interpretable latent variables and its algorithmic implementation is optimized for sparse systems. In particular, it allows for automatic  extraction of main patterns of interactions involving subsets of nodes. We gave examples of this by considering networks of friendship nominations among high school students and of aggressive interactions between monk parakeets. In the case of friendship nominations, \xor highlighted a small subnetwork of four individuals whose interactions stood out from the crowd. Similarly, for the aggression in parakeets, it spotted an individual outlier whose interactions do not seem to align with those observed involving  other individuals in the group. When applied to larger datasets, as the network of hyperlinks between political blogs, the model is still able to identify a dominant mechanism involving a large majority of network nodes.

We considered here an efficient, but possibly limited, mean-field approximation to perform parameters' inference. Its connections with well-known models from statistical physics suggest as natural direction for future developments that of deploying more complex approximations, e.g. using belief propagation \cite{mezard2009information}. While we expect this to lead to more accurate approximate posterior distributions, this may come at the price of increasing complexity, we leave this as an open problem for future work.
From a modelling perspective, an interesting direction for future work is to explore different ways of modelling interaction preferences. Here we assigned a latent variable $\s$ to each node, but it would be interesting to investigate how results change when considering latent variables on edges instead. This choice may be more natural in scenarios where individuals form ties on a case-by-case basis rather than predominantly via one of the two mechanisms explored here. This could potentially account for a further mechanism for edge formation,  as reciprocity \cite{Safdari2020AGM,safdari2021reciprocity,contisciani2021community}. Similarly, when node attributes are available along with the network dataset, it would be compelling to adapt the model to suitably incorporate this extra information using insights from previous works \cite{contisciani2020community}.
Finally, algorithmic developments to ameliorate further runtime and scalability offer another promising paths for future research directions.

In summary, in this work we make a first step to tackle problems with mixed underlying mechanisms determining edge formation in networks. While we showed examples of interesting patterns possibly arising as inferred by our model, we provide an open-source implementation of the code to facilitate future data explorations.

\section*{Data availability}
The code used for the analysis and to generate the synthetic data is publicly available and can be found at \url{https://github.com/liacov/XOR-rankcom}.

\section*{Acknowledgements}
L.I. work was fully supported by the Max Planck Institute for Intelligent Systems. Her current affiliation is Bosch Industry on Campus Lab, University of Tübingen.
\bibliographystyle{apsrev4-1}
\bibliography{bibliography}

\newcommand{\beginsupplement}{%
        \setcounter{table}{0}
        \renewcommand{\thetable}{S\arabic{table}}%
        \setcounter{figure}{0}
        \renewcommand{\thefigure}{S\arabic{figure}}%
        \setcounter{equation}{0}
        \renewcommand{\theequation}{S\arabic{equation}}
         \setcounter{section}{0}
        \renewcommand{\thesection}{S\arabic{section}}
 }

\clearpage
\beginsupplement

\section*{{Supporting Information (SI)}}
\section{Controlling sparsity} \label{S1}

In order to control the average degree value given by the ranking and community structures, we can introduce two parameters $c_{SR}, c_{MT}$ s.t. $c_{SR} = c$ and $w_{kh} = c_{MT} \hat w_{kh} \;, \forall k,h$. We assume that the average degree is composed by three contributes, i.e.:

\be
  \langle k \rangle N = \langle k_{SR} \rangle N + \langle k_{MT} \rangle N + \varepsilon N = \mathbb{E}\rup{\sum_{ij} A_{ij}} \;,
\ee
with $\varepsilon$ noise term given by the outgroup interactions. Under the XOR model, we have:

\be
\mathbb{E}\rup{\sum_{ij} A_{ij}} = \sum_{ij} \sum_{\s} P(\delta_{\s_i\s_j})  \mathbb{E}_{A|\s}\rup{A_{ij}}
=  2\mu(1-\mu)\delta_0 N^2 +  \bup{\mu^2 + (1-\mu)^2} \sum_{ij} \mu S_{ij} + (1 - \mu)M_{ij}
\ee
Hence, imposing the equivalence for each term:

\begin{align}
  \varepsilon &=  2\mu(1-\mu)\delta_0 N \;, \\
  c_{SR} &= \f{  \langle k_{SR} \rangle N }{\mu \bup{\mu^2 + (1-\mu)^2} \sum_{ij} e^{-\beta H_{ij}}} \;, \\
  c_{MT} &= \f{  \langle k_{MT} \rangle N }{(1-\mu) \bup{\mu^2 + (1-\mu)^2} \sum_{ijkh} u_{ik} v_{jh}  \hat w_{kh} } \;.
\end{align}
Notice that this formulation assumes that the outgroup interaction parameter $\delta_0$  has to respect the following bound:

\be
  \delta_0 \leq \frac{\langle k \rangle }{2 \mu (1-\mu) N} \;.
\ee

\section{Mapping to an Ising model} \label{S2}

In order to define this mapping, it is convenient to rewrite the posterior distribution of $\s$ such that:

\be
P(\s|A,\theta) \propto \exp(\log P(\s,A|\theta)) =: \exp(- H_A(\s| J,h)))
\ee
for some energy function

\be
H_A(\s| J,h) = - \sum_i h_i(\theta) \, s_i(\s_i) - \sum_{ij} J_{ij}(\theta) \, s_i(\s_i) \,s_j(\s_j) \;, \quad 2\s_i-1 = s_i(\s_i) \in \{-1,1\} \;.
\ee
We recall that:

\be\label{eq:postsigma}
\log P(\s, A | \theta) = \sum_i \log \bup{ \mu^{\s_i}(1-\mu)^{1-\s_i} } \sum_{ij} \, \log \rup{ \pois(A_{ij};S_{ij})^{\s_i\delta_{\s_i\s_j}} \pois(A_{ij};M_{ij})^{(1-\s_i)\delta_{\s_i\s_j}} \, \pois(A_{ij};\delta_0)^{1-\delta_{\s_i\s_j}} } \;.
\ee
From now on, we will use $\bar S_{ij}, \bar M_{ij}, \bar \lambda_0$ to denote the logarithm of the poisson distributions in $A_{ij}$ with means $S_{ij}, M_{ij}, \lambda_0$. In addition, the dependence on $\theta$ and $\s$ willl be avoided, in order to make the notation easier to read.

Consider the first summation appearing in Eq.~(\ref{eq:postsigma}). It is equal to:

\begin{align}
  \sum_i \s_i \log \mu + (1-\s_i) \log (1-\mu) &= \sum_i \f{s_i+1}2 \log \mu + \bup{ 1-\f{s_i+1}2 } \log (1-\mu) \\
  &= \sum_i \f12 \bup{\log \mu - \log(1-\mu)} s_i + \text{const} \;,
\end{align}
where the constant is not relevant since it will be discarded when normalizing the Boltzmann distribution. Hence we have

\be
h^{(1)}_i \equiv \f12 \bup{\log \mu - \log(1-\mu)}
\ee
as the first component of the $h_i$ field. To obtain the $J_{ij}$ and $h^{(2)}_i = h_i - h^{(1)}_i$ fields, we consider the second summation of Eq.~(\ref{eq:postsigma}), which corresponds to $\log P(A|\s, \theta)$:

\begin{align}
  \log P(A|\s, \theta) = & \sum_{ij} \s_i\delta_{\s_i\s_j} \bar S_{ij} + (1-\s_i)\delta_{\s_i\s_j} \bar M_{ij} + (1-\delta_{\s_i\s_j}) \bar \lambda_0\\
  = &\sum_{ij} \bup{\f{s_i+1}{2}}\bup{\f{s_is_j+1}{2}} \bar S_{ij} + \bup{1-\f{s_i+1}{2}}\bup{\f{s_is_j+1}{2}} \bar M_{ij} \\
  &+ \bup{1-\f{s_is_j+1}{2}} \bar \lambda_0 \\
  = & \sum_{ij} s_is_j \f{\bar{S}_{ij}+\bar{M}_{ij}-2\bar{\lambda}_{0}}4 + \sum_i s_i \sum_{j}\f{\bar{S}_{ij}+\bar{S}_{ji}-\bar{M}_{ij}-\bar{M}_{ji}}4 \\
  &+  \f{\lambda_0}2 N^2 + \sum_{ij} \f{\bar{S}_{ij}+\bar{M}_{ij}}4
\end{align}
Now we have:

\begin{align}
  J_{ij} =& \f{\bar{S}_{ij}+\bar{M}_{ij}-2\bar{\lambda}_{0}}{4}\\
  h_{i}^{(2)} =& \f{1}{4}\sum_{j}(\bar{S}_{ij}+\bar{S}_{ji}-\bar{M}_{ij}-\bar{M}_{ji})
\end{align}
The final Hamiltonian is:

\be
  H_A(s|J,h) = \sum_{i,j}J_{ij}s_{i}s_{j}+\sum_{i}(h_{i}^{(1)}+h_{i}^{(2)})s_i
\ee
where the $J_{ij}$ are asymmetric.

\section{Mean-field approximation for $q$} \label{S3}

We decribe here the procedure followed for computing the approximation of the variational distribution $q$ under mean-field assumption. Firstly, we assume that we can factorize it as $ q(\s) = \prod_i q_i(\s_i), \ q_i(\s_i) = \bern(Q_i)$. Hence, the collection of means $\{Q_i\}_i$ fully charecterises the distribution $q$. We notice that we can write our objective function as

\begin{align}
  \L(q,\theta) = & \; Q_i \; \mathbb{E}_{j\neq i}\big[ \log P(A,\s_i = 1, \s_{j\neq i}|\theta) \big]
  + (1 - Q_i) \; \mathbb{E}_{j\neq i}\big[ \log P(A,\s_i = 0, \s_{j\neq i}|\theta) \big] + H_b(Q_i) + \text{const} \;.
\end{align}
Maximizing w.r.t. $Q_i$ gives the equation:

\be
  \log \bup{\f{Q_i}{1-Q_i}} = \mathbb{E}_{j\neq i}\big[ \log P(A,\s_i = 1, \s_{j\neq i}|\theta) \big]  -  \mathbb{E}_{j\neq i}\big[ \log P(A,\s_i = 0, \s_{j\neq i}|\theta) \big]\;,
\ee
which is solved for

\begin{align}
  Q_i &= \f{ \exp \bup{\mathbb{E}_{j\neq i}\big[ \log P(A,\s_i = 1, \s_{j\neq i}|\theta) \big]  -  \mathbb{E}_{j\neq i}\big[ \log P(A,\s_i = 0, \s_{j\neq i}|\theta) \big]}}{1 + \exp \bup{\mathbb{E}_{j\neq i}\big[ \log P(A,\s_i = 1, \s_{j\neq i}|\theta) \big]  -  \mathbb{E}_{j\neq i}\big[ \log P(A,\s_i = 0, \s_{j\neq i}|\theta) \big]}} \\
  &= \f{ \exp \bup{\mathbb{E}_{j\neq i}\big[ \log P(A,\s_i = 1, \s_{j\neq i}|\theta) \big] }}{ \exp \bup{\mathbb{E}_{j\neq i}\big[ \log P(A,\s_i = 1, \s_{j\neq i}|\theta) \big]  +  \mathbb{E}_{j\neq i}\big[ \log P(A,\s_i = 0, \s_{j\neq i}|\theta)}}\\
  &=  \f{ \exp \bup{\mathbb{E}_{j\neq i}\big[ \log P(\s_i = 1|A, \theta, \s_{j\neq i}) \big] }}{ \exp \bup{\mathbb{E}_{j\neq i}\big[ \log P(\s_i = 1,|A, \theta, \s_{j\neq i}) \big]  +  \mathbb{E}_{j\neq i}\big[ \log P(\s_i = 0|A, \theta, \s_{j\neq i})}}\;.
\end{align}
Since the formula for $P(A,\s|\theta)$ is known, we can compute this quantity and obtain $Q_i = q_i(\s_i = 1)$. Using the $\delta_{\s_l\s_k}$ parametrization, neglecting self-loops and discarding the terms not dependent on $\s_i$, we finally obtain the following formula:

\begin{align}
  Q_{i} &= \f{f_{i1}}{f_{i1} + f_{i2}} \;, \label{eq:updateQ} \\ 
  f_{i1} &= \mu \prod_{j\neq i}\rup{\pois(A_{ij};S_{ij})\pois(A_{ji};S_{ji})}^{Q^{}_{j}}\rup{\pois(A_{ij};\lambda_{0})\pois(A_{ji};\lambda_{0})}^{(1-2Q^{}_{j})} \;, \label{eq:updateQ1} \\
  f_{i2} &= (1-\mu) \prod_{j\neq i}\rup{\pois(A_{ij};M_{ij})\pois(A_{ji};M_{ji})}^{Q^{}_{j}-1} \label{eq:updateQ2} \;.
\end{align}

\section{$\L$ maximization under mean-field assumption} \label{S4}

Here we report the computations for defining the EM algorithm updates. The goal is to maximise the following quantity:

\begin{align} \label{eq:explL-n}
  \L(q,\theta) = \sum_{\s} q(\s)\, \log \f{P(\s,\theta| A)}{q(\s)} = \sum_{\s} q(\s)\, \log P(A |  \s, \theta) +  \sum_{\s} q(\s) \, \log P(\s|\mu) + H(q)
\end{align}
where $H$ is the Entropy function. As a first step, we can write the log-likelihood term more explicitly as
\begin{align}\label{eq:qlikA}
  \begin{split}
    \sum_{\s} q(\s)\,\log P(A |  \s, \theta) = \sum_{i,j,\s} & \bigg[ \s_i q(\s) \, \delta_{\s_i\s_j} \log\pois(A_{ij};S_{ij}) + (1-\s_i ) \,q(\s)\,  \delta_{\s_i\s_j} \log\pois(A_{ij};M_{ij}) \\
    & + q(\s)(1- \delta_{\s_i\s_j})\, \log\pois(A_{ij};\delta_0) \bigg]\;.
  \end{split}
\end{align}
Now consider the first term in Eq.~(\ref{eq:qlikA}).
\begin{align}
  \sum_{i,j,\s} \s_i q(\s) \, \delta_{\s_i\s_j} \log\pois(A_{ij};S_{ij}) &= \sum_{ij} Y_{ij}  \log\pois(A_{ij};S_{ij})\;;  \\
   Y_{ij} &:= \sum_{\s} \s_i q(\s)\, \delta_{\s_i\s_j} \;.
\end{align}
Since we work under the MF assumption, the following equality holds:
\be\label{eq:Yij}
 Y_{ij} = \sum_{\s} \s_i q(\s)\, \delta_{\s_i\s_j} = \sum_{\s}q(\s)\bup{2\s_{i}^{2}\s_{j}-\s_{i}^{2}-\s_{i}\s_{j}+\s_{i}} = \; \Exp\rup{\s_{i}} \Exp\rup{\s_{j}} = Q_iQ_j\;.
\ee
As for the second term in Eq.~(\ref{eq:qlikA}), we notice that for similar reasons the sum on $\s$ can be rewritten as
\begin{align}
\sum_{\s} (1-\s_i ) \,q(\s)\,  \delta_{\s_i\s_j} &= \sum_{\s}  \,q(\s)\,  \delta_{\s_i\s_j} - \s_i \,q(\s)\,  \delta_{\s_i\s_j} = X_{ij} - Y_{ij} \;, \\
X_{ij} &:=  \sum_{\s}  \,q(\s)\,  \delta_{\s_i\s_j} = 2Q_iQ_j -Q_i - Q_j +1\;.
\end{align}
Plugging these results into the log-likelihood term, we obtain:
\begin{align}\label{eq:qlikA2}
  \begin{split}
    \sum_{\s} q(\s)\, \log P(A |  \s, \theta) = \sum_{ij} \, \bigg[ Y_{ij} \log\pois(A_{ij};S_{ij}) + (X_{ij}-Y_{ij}) \log\pois(A_{ij};M_{ij}) +(1-X_{ij}) \log\pois(A_{ij};\delta_0) \bigg]\;,
  \end{split}
\end{align}
and the whole $\L$ in Eq.~(\ref{eq:explL-n}) now is
\begin{align} \label{eq:finalL}
  \L(q,\theta) = &\sum_{ij} \, Y_{ij} (-S_{ij}+ A_{ij}\log S_{ij}) + (X_{ij}-Y_{ij}) (-M_{ij}+ A_{ij}\log M_{ij}) + \sum_{ij}(1-X_{ij}) (-\delta_0+ A_{ij}\log \delta_0) \\
  &+ \sum_i Q_i\log(\mu) + (1-Q_i)\log(1-\mu)  + H(q)  \\
  =& \sum_{ij} \, Q_iQ_j (-S_{ij}+ A_{ij}\log S_{ij}) + \sum_{ij} \, (Q_iQ_j -Q_i - Q_j +1) (-M_{ij}+ A_{ij}\log M_{ij})\\
  &- \sum_{ij}\, (2Q_iQ_j -Q_i - Q_j) (-\delta_0+ A_{ij}\log \delta_0) + \sum_i Q_i\log(\mu) + (1-Q_i)\log(1-\mu)  + \sum_i H_b(Q_i) \;.
\end{align}

We now have all the elements for writing the EM algorithm updates. However, some additional steps are needed for defining computationally efficient equations.

In order to have write the community parameters in such a way that at the Maximization step each $u_{ik}$ and $v_{jh}$ is independent by all the other elements of the vectors $u_{i}$ and $v_{j}$, we introduce the variational probability:
\be
\rho_{ijkh}=\f{u_{ik}v_{jh}w_{kh}}{\sum_{kh}u_{ik}v_{jh}w_{kh}} \;.
\ee
that is representing the probability of observing an interaction between the nodes $i,j$ because of their respective belonging to communities $k,h$. It will be evaluated in Expectation phase. Deriving the $\L$ w.r.t. $u_{ik}$ we obtain:
\be
u_{ik} = \f{\sum_{jh} (Q_iQ_j -Q_i - Q_j +1)\, A_{ij}\rho_{ijkh} }{\sum_{jh}(Q_iQ_j -Q_i - Q_j +1)\, v_{jh}w_{kh}} \;.
\ee
We find similar expression for $v_{ih}$ and $w_{kh}$. For the remaining quantities $s_i, \mu, c$ and $\delta_0$, maximizing the $\L$ leads to the updates:
\begin{align}
s_i &= \f{\sum_{j} Q_{i}Q_{j} \,s_j \rup{A_{ji} + A_{ij}  } + Q_{i}Q_{j} \rup{ A_{ij} - A_{ji} }}{\sum_j Q_{i}Q_{j} \rup{A_{ji} + A_{ij}}} \;, \quad \mu = \f1N\sum_{i} Q_{i} \;, \\
c &= \f{\sum_{ij} Q_{i}Q_{j}\, A_{ij}}{\sum_{ij}Q_{i}Q_{j} e^{-\f{\beta}{2} (s_{i}-s_{j}-1)^{2}}} \;, \quad \delta_0 = \f{\sum_{ij}A_{ij} (2Q_iQ_j -Q_i - Q_j) }{\sum_{ij} (2Q_iQ_j -Q_i - Q_j) } \;.
\end{align}


\end{document}